\documentclass[prd, nofootinbib, notitlepage, twocolumn, superscriptaddress]{revtex4-1}

\usepackage{graphicx}
\usepackage{amsmath}
\usepackage{ulem}
\usepackage{url}

\usepackage{xcolor}
\definecolor{naviBlue}{RGB}{0,0,128}
\usepackage[colorlinks  = true,
            linkcolor   = naviBlue,
            urlcolor    = naviBlue,
            citecolor   = naviBlue,
            anchorcolor = naviBlue]{hyperref}

\newcommand{\gsim}{\lower.7ex\hbox{$\;\stackrel{\textstyle>}{\sim}\;$}}
\newcommand{\lsim}{\lower.7ex\hbox{$\;\stackrel{\textstyle<}{\sim}\;$}}            

\newcommand{\appref} [1]{\hyperref[app::#1]{Appendix~\ref*{app::#1}}}
\newcommand{\secref} [1]{\hyperref[sec::#1]{Sec.~\ref*{sec::#1}}}
\newcommand{\figref} [1]{\hyperref[fig::#1]{Fig.~\ref*{fig::#1}}}
\newcommand{\tabref} [1]{\hyperref[tab::#1]{Table~\ref*{tab::#1}}}
\newcommand{\eqnref} [1]{\hyperref[eqn::#1]{Eq.~(\ref*{eqn::#1})}}

\newcommand{\etal}{{\it et al.}}
\newcommand{\etalS}{{\it et al. }}

\newcommand{\pbar}{\bar{p}}
\newcommand{\Tp}{T_p}
\newcommand{\Tproj}{T_i}
\newcommand{\pb}{{\bar{p}}}

\newcommand{\Tpbar}{T_{\bar{p}}}
\newcommand{\Epbar}{E_{\bar{p}}}
\newcommand{\sS}{\sqrt{s}}
\newcommand{\xR}{x_\mathrm{R}}
\newcommand{\xF}{x_f}
\newcommand{\pT}{p_{\mathrm{T}}}
\newcommand{\pL}{p_{\mathrm{L}}}

\newcommand{\sigmaInv}{\sigma_{\mathrm{inv}}}

\begin{document}

\title{Production cross sections of cosmic antiprotons in the light of new data from the NA61 and LHCb experiments}

\author{Michael Korsmeier}
\email{korsmeier@physik.rwth-aachen.de}
\affiliation{Dipartimento di Fisica, Universit\`a di Torino, via P. Giuria 1, 10125 Torino, Italy}
\affiliation{Istituto Nazionale di Fisica Nucleare, Sezione di Torino, Via P. Giuria 1, 10125 Torino, Italy}
\affiliation{Institute for Theoretical Particle Physics and Cosmology, RWTH Aachen University, 52056 Aachen, Germany}
\author{Fiorenza Donato}
\email{donato@to.infn.it}
\affiliation{Dipartimento di Fisica, Universit\`a di Torino, via P. Giuria 1, 10125 Torino, Italy}
\affiliation{Istituto Nazionale di Fisica Nucleare, Sezione di Torino, Via P. Giuria 1, 10125 Torino, Italy}
\author{Mattia Di Mauro}
\email{mdimauro@slac.stanford.edu}
\affiliation{W.W. Hansen Experimental Physics Laboratory, Kavli Institute for Particle Astrophysics and Cosmology, 
              Department of Physics and SLAC National Accelerator Laboratory, Stanford University, Stanford, California 94305, USA}

\begin{abstract}

\noindent
The cosmic-ray flux of antiprotons is measured with high precision by the space-borne particle spectrometers AMS-02.
Its interpretation requires a correct description of the dominant production process for antiprotons in our Galaxy, namely, 
 the interaction of cosmic-ray proton and helium with the interstellar medium.
In light of new cross section measurements by the NA61 experiment of 
$p + p \rightarrow \pbar + X$ 
and the first ever measurement of 
$p + \mathrm{He} \rightarrow \pbar + X$ 
by the LHCb experiment, we update the parametrization of proton-proton and proton-nucleon cross sections. 
\\
We find that the LHCb $p$He data constrain a shape for the cross section at high energies and show for the first time how well the rescaling from the $pp$ channel applies to a helium target. By using $pp$, $p$He and $p$C data we estimate the uncertainty on the 
Lorentz invariant cross section for $p + \mathrm{He} \rightarrow \pbar + X$. 
We use these new cross sections to compute the source term for all the production channels, considering also nuclei heavier than He both in cosmic rays and the interstellar medium. The uncertainties on the total source term are up to $\pm20$\%. and slightly increase below antiproton energies of 5~GeV. 
This uncertainty is dominated by the $p+p \rightarrow \pbar + X$ cross section, which translates into all channels 
since we derive them using the $pp$ cross sections. 
The cross sections to calculate the source spectra from all relevant cosmic-ray isotopes are provided in the Supplemental Material.
We finally quantify the necessity of new data on antiproton production cross sections, and pin down the kinematic parameter space 
which should be covered by future data.
\end{abstract}

\maketitle

\section*{\label{sec::introduction}Introduction}

With the last generation of particle detectors in space, physics of charged cosmic rays (CRs) has become a precision discipline. 
During the last decade, the space-based spectrometers PAMELA and AMS-02, which is borne to the International Space Station, 
have driven measurement uncertainties in the CR fluxes as low as the percent level in an energy range from 1~GeV to a few TeV. 
They have measured  the CR nuclear  
\cite{PAMELA_Adriani:2011cu,2014ApJ...791...93A,AMS-02_Aguilar:2015_ProtonFlux,AMS-02_Aguilar:2015_HeliumFlux,Aguilar:2016vqr} 
and leptonic (positron and electron)  
\cite{2009Natur.458..607A,2011PhRvL.106t1101A,2013PhRvL.110n1102A,2014PhRvL.113l1102A,2014PhRvL.113v1102A} 
components, as well as CR antiprotons 
\cite{2010PhRvL.105l1101A,AMS-02_Aguilar:2016_AntiprotonFlux}. 
The most recent antiproton flux measurement by AMS-02 extends from 1 to 400 GeV with an uncertainty of 5\% for almost the whole energy range.   
The new precise flux data have stimulated various analyses on Galactic CR propagation and particle dark matter 
annihilation into antimatter (see, e.g., \cite{Kappl:2014_pbarCrossSection,2016PhRvD..94l3019K,Cuoco:2017rxb,Winkler:2017xor}).
However, to infer correct conclusions on any modeling and interpretation, an accurate description of the underling antiproton production is necessary. 
It is generally established that the bulk of antiprotons in our Galaxy is produced by the interaction of CRs 
on the interstellar medium (ISM)~\cite{Donato:2008jk}, conventionally called secondary antiprotons.
In practice, the dominant contribution is provided by the proton-proton ($pp$) channel, namely CR proton on ISM hydrogen, 
and either the CR projectile or the ISM target replaced by helium (He$p$, $p$He, and HeHe). Heavier channels can contribute at the few percent level.

In a first paper \cite{Donato:2017ywo} (hereafter DKD17), we discussed the requirement on cross section measurements 
to determine the antiproton source term at the uncertainty level of AMS-02 flux data. 
Here we invert our perspective and seek to determine the source term and its uncertainty from existing cross section measurements. 
In general, there are two different strategies to parametrize the energy-differential cross section which enters the source term calculation.
The first possibility is to find an analytic description of the fully-differential and Lorentz invariant cross section performing a fit to cross section data. Then, a Lorentz transformation and angular integration are applied to find the energy-differential cross section. 
This strategy was first pursued by~\cite{TanNg:1983_pbarCrossSectionParametrization}. 
The other option is to use Monte Carlo predictions to extract the required cross section.
This was, for example, done with DTUNUC \cite{Simon_Antiproton_CS_Scaling_1998, Donato:2001_DTUNUC} 
and more recently using EPOS-LHC and QGSJET-II-04 by Kachelriess \etalS~\cite{Kachelriess:2015_pbarCrossSection}  (hereafter KMO). 
These Monte Carlos, EPOS and QGSJET, were originally developed for very high-energy interactions as for example occurring in CR air showers, 
but a special training to low-energy data allows to apply them to Galactic CR antiprotons.
However, this approach only gives robust results above an antiproton energy of 10~GeV, 
which is a drawback considering that AMS-02 antiproton data reach down to 1~GeV. 
Analytic parameterizations overcome this limit,  since they are directly based on the relevant low-energy data. 
We pursue an analytic parametrization in the following. 
With respect to previous analysis \cite{diMauro:2014_pbarCrossSectionParametrization, Kappl:2014_pbarCrossSection, Winkler:2017xor}, 
we exploit new data by the NA61 experiment~\cite{Aduszkiewicz:2017sei} in the $pp$ channel and the first ever determination of $p$He data by the LHCb experiment~\cite{Graziani:2017}.

\section{\label{sec::current_status} State of the art  on the antiproton source term }

We first shortly review the basic equations and procedure to calculate the cosmic antiproton source term.  
The source term $q_{ij}$ originating from the interaction of the CR component $i$ on the ISM component $j$ is given by a convolution integral 
of the energy-differential antiproton production cross section $d\sigma_{ij}/d\Tpbar$ with the incoming CR flux $\phi_i$ 
and the ISM target density $n_{\mathrm{ISM},j}$ over the CR kinetic energy per nucleon $T_i$:
\begin{eqnarray}
  \label{eqn::sourceTerm_1}
  q_{ij}(\Tpbar) &=& \int\limits_{T_{\rm th}}^\infty d\Tproj \,\, 
                                    4\pi \,n_{\mathrm{ISM},j} \, \phi_i  (\Tproj) \, \frac{d\sigma_{ij}}{d \Tpbar}(\Tproj , \Tpbar).
\end{eqnarray}
Here $T_{\rm th}$ is the energy threshold for antiproton production and the factor $4\pi$ corresponds the angular integration of an isotropic CR flux. 
However, experiments do not directly measure the energy-differential cross section 
but rather the fully-differential cross section, usually expressed in a Lorentz invariant form 
\begin{eqnarray}
  \label{eqn::inv_XS}
  \sigmaInv = E \frac{d^3\sigma}{dp^3} (\sS,\xR,\pT),
\end{eqnarray}
where $E$ is the total antiproton energy and $p$ its momentum. 
It is typically a function of the kinematic variables 
$\sS$, $\xR=\Epbar^*/\Epbar^\mathrm{max\,*}$, $\pT$,
which are the center-of-mass (CM) energy, the antiproton energy divided by the maximal antiproton energy in the CM frame, 
and the transverse antiproton momentum, respectively\footnote{The superscript * denotes that a quantity is taken in the CM frame.}.
Notice that the kinematic variables always refer 
to the nucleon-nucleon CM frame. Sometimes it is convenient to replace $\xR$ by the so-called Feynman variable 
$\xF=2\pL^*/\sS$
which is twice the longitudinal antiproton momentum divided by the 
CM energy. 
To obtain the energy-differential cross section in \eqnref{sourceTerm_1}, 
the kinetic variables are transferred into the LAB frame, {\it i.e.} the frame where the target particle is at rest, by means of a Lorentz transformation. 
Suitable variables in the LAB frame are the kinetic energies of the CR $T_i$ and of
the antiproton $\Tpbar$, and the angle $\theta$ of the produced antiproton with respect to the incident CR. 
Finally, the integral over the solid angle $\Omega$ relates to the energy-differential cross sections:
\begin{eqnarray}
\label{eqn::energyDifferentialToInvCS}
\frac{d\sigma_{ij}}{d \Tpbar}(T , \Tpbar) 
  &=&  p_\pb  \int d \Omega \,\, \sigma_\mathrm{inv}^{(ij)}(T_i, \Tpbar, \theta).
\end{eqnarray}
More details and explicit calculations are presented in DKD17. 

The dominant channels for the production of 
secondary antiprotons are $pp$ at roughly 50-60\% of the total spectrum, and $p$He and He$p$ at 10-20\% each, while the channels 
involving heavier incoming CRs contribute only up to a few percent (see below). 
Until very recently, no measurements of the helium channels were available, rendering the $pp$ channel the baseline 
for any scaling to proton-nucleus ($pA$) or nucleus-nucleus ($AA$) channels. 
Often this was done in a very simplified  assumption, where each nucleon of the target interacts with each nucleon of the projectile, such that 
the nucleon-nucleon interaction scales according to the $pp$ one. To first order, one expects a re-scaling of $pp$ by a factor $A^D$. 
The value of the parameter $D$ is typically chosen between $2/3$, 
corresponding to an approximation in which the target area is estimated from a classical sphere, and $1$, if all nucleons interact completely independently.
A parameterization for the cross sections involving helium (as projectile and/or target) should be derived phenomenologically 
from data in this specific channel. In order to be reliable, the data should cover a wide portion of the kinematical parameter space 
relevant for antiproton energies of interest. As we will discuss below, the first data on $p$He scattering do not fulfil this condition. 
Therefore, in the following of this paper we  rely on re-scaling from $pp$ cross sections, 
which implies that uncertainties from this channel almost directly translate into the helium and heavier channels. 
Given the importance of the $p+p\rightarrow\bar{p} + X$ channel, we review it below.   

\subsection{The proton-proton channel}

The latest analyses of the antiproton production cross section were done by 
Di Mauro \etalS~\cite{diMauro:2014_pbarCrossSectionParametrization} and 
Kappl \etalS~\cite{Kappl:2014_pbarCrossSection}, which were both triggered by newly available precise data 
from the NA49 experiment \cite{NA49_Anticic:2010_ppCollision}. NA49 measures antiprotons in a fix-target $pp$ collision at $\sS=17.3$~GeV. 
The two analyses follow slightly different strategies. 
On the one hand,  \cite{diMauro:2014_pbarCrossSectionParametrization} combined the NA49 data with a series of old data sets
spanning a range in CM energies from $\sS=6.1$~GeV to $200$~GeV. 
They perform a global fit to extract a parametrization of the fully-differential Lorentz invariant cross section. 
On the other hand, \cite{Kappl:2014_pbarCrossSection} relies on NA49 data and exploits the scaling invariance of the cross section above $\sqrt{s} =$10~GeV, 
namely the fact that the cross section does not depend on $\sS$: $\sigmaInv (\sS, \xR, \pT) \approx \sigmaInv(\xR, \pT)$. 
Below 10~GeV, the scaling invariance is supposed to be violated. 
The two analyses pointed out  two issues not  considered in previous parameterizations: isospin violation and hyperon induced production.
In order to calculate the total amount of antiprotons produced in our Galaxy, one has to include all the particles which decay into antiprotons,
namely antineutrons and antihyperons.  
Traditionally,  it has been assumed that antiproton and antineutron production in $pp$ collisions is equal, and the 
antiproton source term has simply been multiplied by a factor 2 to account for the contribution from antineutron decays. 
Indeed, NA49 data \cite{Fischer:2003xh} indicate an enhanced production of antineutrons with respect to the antiproton one. 
Following \cite{Winkler:2017xor}, we consider a $\sS$ dependent  isospin violation, which is estimated not to exceed  20\%. 
The second issue has a similar origin. A fraction of the total antiproton yield originates from an intermediate antihyperon, 
which subsequently decays to an antiproton. The NA49 collaboration 
explicitly corrects and subtracts antiprotons originating from hyperons. However, 
the hyperon correction in older experiments is not always clearly taken into account, and data are not easily comparable. 
The usual assumption is that those experiments  were not able to distinguish between primary (prompt) antiprotons and intermediate hyperon states,  and contain a hyperon contamination  which is of the order of 30\%-60\%.  
In an update of  \cite{Kappl:2014_pbarCrossSection}, Winkler \cite{Winkler:2017xor} discusses the energy dependence of isospin violation and hyperon production. 
Furthermore, he points out  that the scaling invariance of the cross section is broken above $\sS=50$~GeV 
such that the $\pT$-shape and normalization of the cross section require to be adjusted. High-energy collider data 
are used to specify and parametrize the scaling violation. 
Finally, above $\sS$=10~GeV the analytic result in \cite{Winkler:2017xor} agrees with the Monte Carlo approach 
by KMO, hinting that towards high energies the descriptions become robust, which is expected since the cross sections are constrained by precise NA49 and LHC data.
Below 10~GeV the situation is different, because the relevant data taken in the 70's or 80's incorporate large (systematic) uncertainties.

\begin{figure}[t!]
	\includegraphics[width=1.\linewidth]
	    {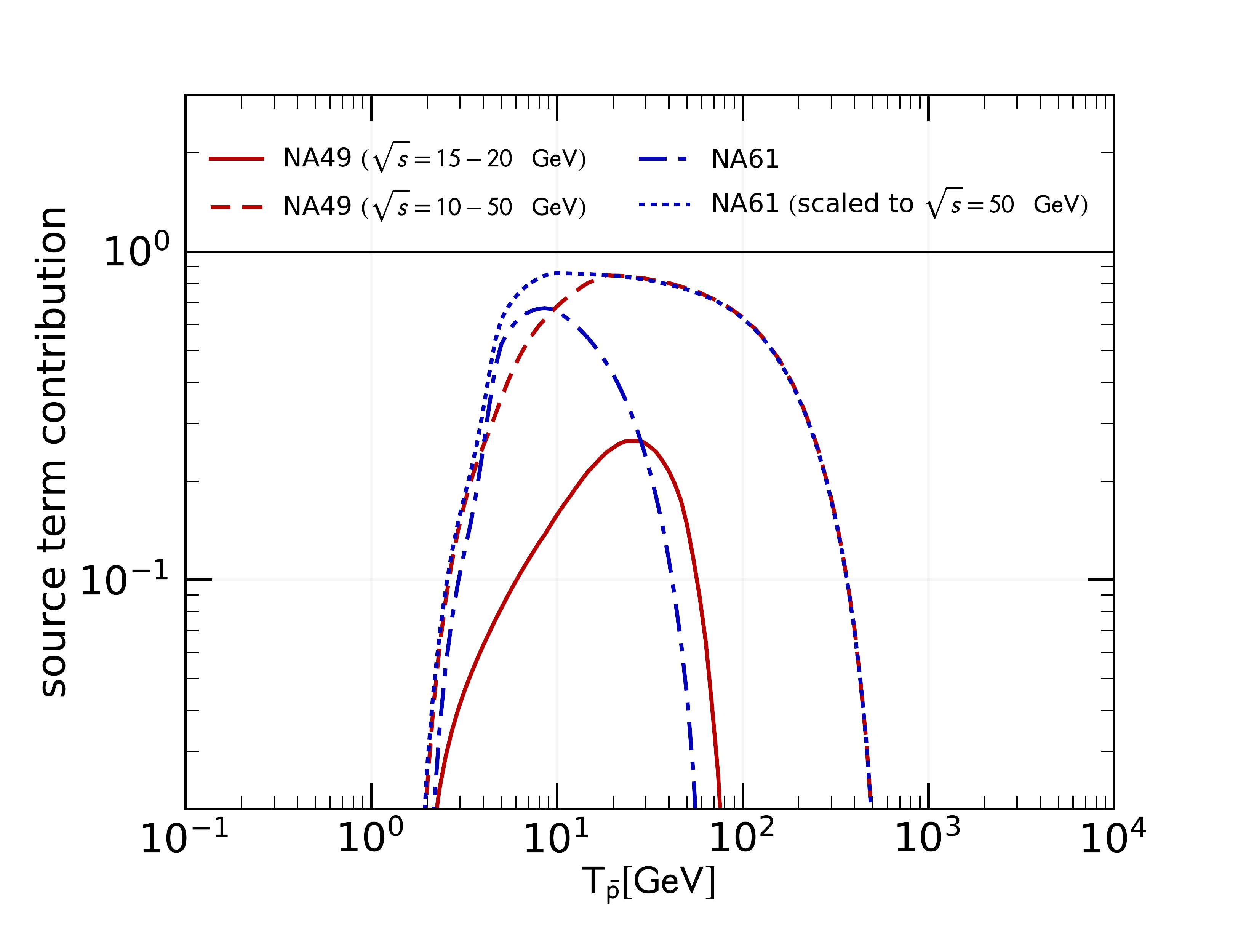}
  \caption{ Fraction of the $pp$ source term originating from the kinematic parameter space of the cross section 
            which is experimentally determined by NA49 and NA61.
            The contribution is normalized to the total $pp$ source term. The NA61 data are taken for $\sS=7.7$~GeV to $17.3$~GeV (blue dot-dashed line), 
            while the NA49 is taken at $\sS=17.3$~GeV and here assumed to be valid 
            in the range 15-20~GeV (solid red line). The red dashed line is obtained assuming that the NA49 data are valid 
            in the $\sS$ range from 10 to 50~GeV, while the dotted blue one is obtained extending the validity of NA61 data  
            up to $\sS=50$~GeV. More explicitly, we assume a range from $\sS=7.7$~GeV to $50$~GeV.} 
	\label{fig::contribution_pp}
\end{figure}

Very recently the NA61 experiment published antiproton cross section measurements at four different CM energies 
$\sS$=7.7, 8.8, 12.3 and 17.3~GeV, corresponding to beam proton energies $T_p$=31, 40, 80, and 158 GeV, respectively \cite{Aduszkiewicz:2017sei}. 
The data are corrected for hyperons and, compared to NA49, extend to lower $\sS$. 
To see how much the NA61 data improve our knowledge about the $pp$ antiproton source term, we conduct the following exercise. 
We calculate the fraction of the $pp$ source term originating from the kinematic parameter space of the cross section 
which is experimentally determined by NA49 and NA61, respectively.
\figref{contribution_pp} shows this fraction normalized to the total $pp$ source term, $i.e.$ integrated on the whole kinematic parameter space.
In more detail, the source term in \eqnref{sourceTerm_1} contains an integral over $\Tp$, or equivalently $\sS$, 
while NA49 data are taken for one fixed value of $\sS$. 
In order to extract meaningful results we have to know the cross section over a non-zero range in $\sS$.
A conservative assumption is that the NA49 cross section is known in a small range around 17.3 GeV, we choose $\sS=15$ to 20~GeV. 
From \figref{contribution_pp} we draw the conclusion
that the experimental data of NA49 (narrow $\sS$ range) contribute 20\% to the antiproton source spectrum,
peak around $\Tpbar=30$~GeV, and quickly decrease towards smaller or larger energies. 
The information contained in this data gets totally negligible for $\Tpbar < 15$ GeV and $\Tpbar > 70$ GeV. 
In contrast to NA49, the NA61 experiment performed runs also at lower $\sS$, which significantly
improves the coverage of the contribution to the source spectrum. 
The  experimental data of NA61 account for up to 70\% and peak at $\Tpbar$ around 8~GeV.
As a matter of fact, the contribution of the true experimental data to the total source spectrum 
covers a relatively small range in$\Tpbar$. One might wonder how this 
can lead to an accurate determination of the source term spectrum. The reason is the 
theoretical assumption of scaling invariance, according to which  the cross section is independent of $\sS$ in a range from 10 to 50~GeV~\cite{Winkler:2017xor}. 
In other words, we can pretend to know the cross section from $\sS=10$ to 50~GeV from a single measurement within the range.
We therefore extend the validity of both the experiments accordingly. The results in \figref{contribution_pp}  show that 
the NA49 parameter space can contribute between 70\% and 80\% from $\Tpbar\sim10$ to 100~GeV.
Above this energy, the determination of the source spectrum requires further data at large $\sS$ describing the scaling violation.
The extended NA61 data coincide with NA49 above $\Tpbar\sim20$~GeV, while significantly 
improving the coverage of the source spectrum at lower energies down to 5~GeV.
The baseline for our calculation in \figref{contribution_pp}  is the cross section parametrization 
derived later in this paper (Param.~II-B). 
However, the results are expected to be robust against changing the actual parametrization.

The conclusion of this exercise is that, in order to constrain the $pp$ source term for $\Tpbar\lsim 5$~GeV, 
it is necessary to have additional low-energy data available. 
Indeed, the currently available cross section measurements below $\sS\approx7$~GeV  contain large systematic uncertainties, 
such that a good determination is hard to obtain. 
We notice that it would be useful to collect precise data at low $\sS$ to fix the antiproton spectrum in all the energy range 
where CR data are now provided with an extremely high accuracy \cite{AMS-02_Aguilar:2016_AntiprotonFlux}. 
Especially, progress could be made by a $p+p\rightarrow \pbar +X$
experiment at energies below  $\sS=7$~GeV.
In \appref{app_source_term_fraction} we show how data from NA61 at $\sS=6.3$~GeV could improve the cross section coverage of the $pp$ source term.
A detailed study of the complete relevant parameter space is discussed in DKD17.

\subsection{The nuclear channels}

In addition to the production of antiprotons from $pp$ scatterings,  
the $p$He and He$p$ channels contribute a large fraction of  the total source term. 
\begin{figure}[b!]
	\includegraphics[width=1.\linewidth]
	    {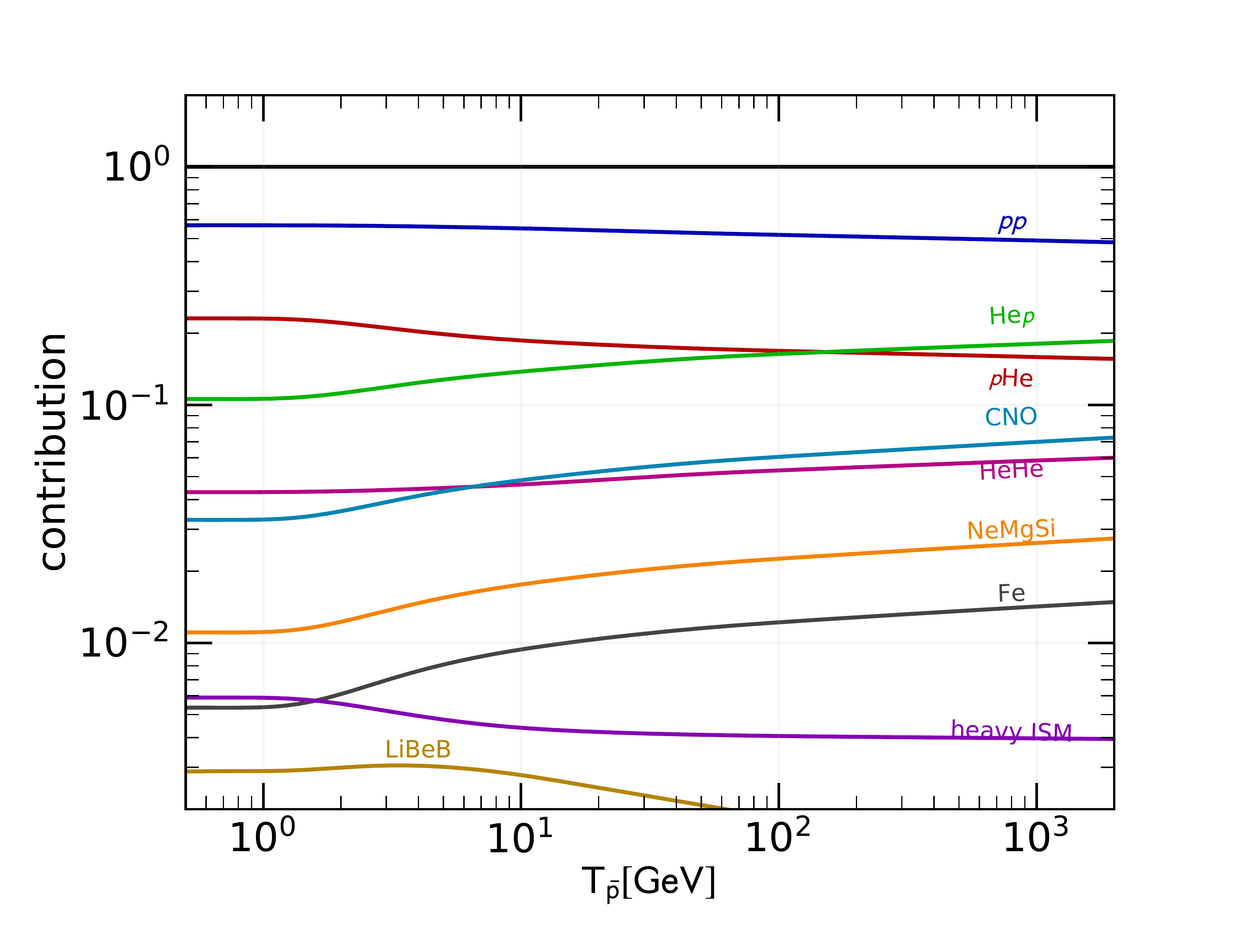}
  \caption{ Relative contribution of the various production channels to the total secondary antiproton source spectrum. The four dominant channels 
            $pp$, $p$He, He$p$, and HeHe are given individually. We group heavy CR nuclei scattering off hydrogen and helium in the ISM:
            CNO, NeMgSi, Fe, and LiBeB. By heavy ISM we denote CR proton and helium scattering off the rare ISM components CNONeMgSiFe. 
           }
	\label{fig::contribution_sourceTerm}
\end{figure}
This information may be inferred from \figref{contribution_sourceTerm}, where we plot the relative contribution of each 
production channel obtained by changing the incoming CR nuclei and the ISM components. 
The production cross sections are taken from the results we present in \secref{pA_fit} (Param. II-B).
In the figure, $pp$, $p$He, He$p$, HeHe label the CR-ISM nucleus. For heavier CR nuclei, we group the 
reactions of LiBeB, CNO, Fe and NeMgSi CR nuclei over the ISM ($p$ and He). We also consider the contribution 
from CR $p$ and He scattering off  the subdominant heavy ISM components accounted for the  CNONeMgSiFe nuclei.
The CR fluxes have been taken as follows:
$p$ from \cite{AMS-02_Aguilar:2015_ProtonFlux}, He from \cite{AMS-02_Aguilar:2015_HeliumFlux}, 
Li, Be and B from \cite{Aguilar:2018njt}, C and O from~\cite{Aguilar:2017hno}, N from~\cite{Yan:2017_TalkXSCRC}, while for all the heavier nuclei we apply the rescaling to oxygen flux as in \cite{1990AandA...233...96E}.
For the ISM composition, we assume $n_{\rm H} = 1 \; {\rm cm}^{-3}$, $n_{\rm He} = 0.1\, n_{\rm H} $, while the abundance for heavier nuclei 
is taken from \cite{Przybilla:2008ph}. 
It is clear from the figure that the channel involving He, both projectile and target, constitute 30-40\%  of the total 
spectrum depending on the antiproton energy. 
The heavier primary CNO nuclei contribute a non negligible few percent at the AMS energies. 
All the other contributions considered in this study turn out to be negligible. 

Until very recently the  cross sections involving He nuclei were not experimentally determined, and  
all calculations rely on re-scaling and extrapolation from $pp$ and $pA$ measurements, 
where $A$ is typically carbon, but sometimes  heavier nuclei up to lead.
The strategy for re-scaling was either based on Monte Carlo simulations, as performed with DTUNUC at low energies 
\cite{Donato:2001_DTUNUC} or KMO at high energies, 
or on fitting parameterizations to the scarce $pA$ data, as performed by Duperray \etalS
\cite{Duperray:2003_pbarCrossSectionParametrizationForPA}. 
The LHCb collaboration provides now the first ever measurement of  $p+\mathrm{He} \rightarrow \pb + X$~\cite{Graziani:2017}, where the incident 
LHC protons of 6.5~TeV momentum scatter off a fixed-target helium (corresponding to $\sS = 110$~GeV). 
The LHCb detector can measure antiprotons with a momentum between 10 and 100 GeV and transverse momentum varying between 0.5 and 3.4 GeV.
In~\cite{Reinert:2017aga} these data are compared to the parametirization of~\cite{Winkler:2017xor} showing reasonable agreement.
\begin{figure}[b!]
	\includegraphics[width=1.\linewidth]
	    {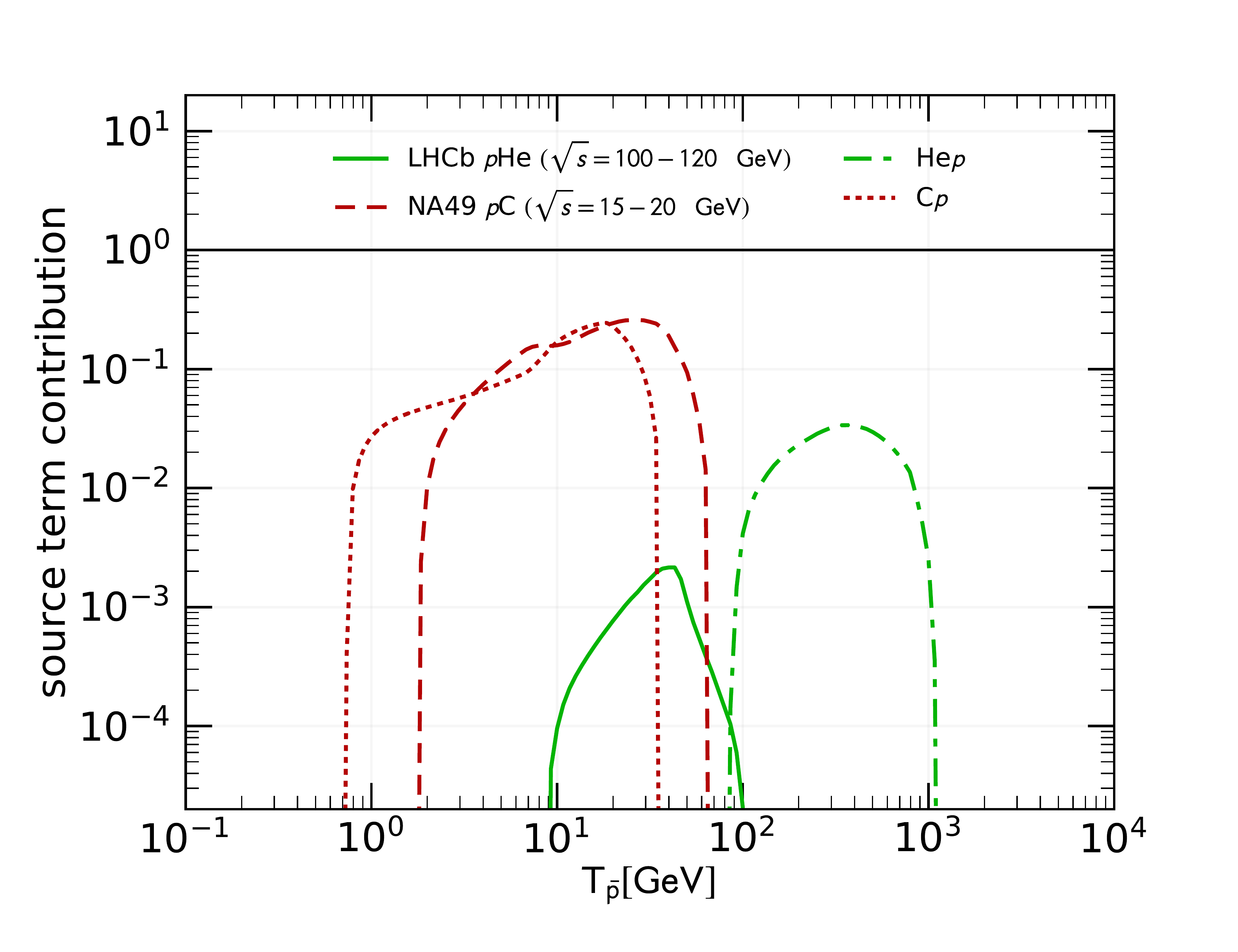}
  \caption{ Similar to \figref{contribution_pp}, but for the nuclear channel.             
            It shows the fraction of the antiproton source term which is 
            covered by the kinematic parameter space of the cross section 
            measurements by NA49 $p$C and LHCb $p$He. 
            Specifically, we assume a range of $\sS=15$-20 GeV for NA49 and 100-120 GeV for LHCb.
            Each contribution is normalized to the total source term of the specific  channel. }
	\label{fig::contribution_pHe}
\end{figure}
\figref{contribution_pHe} shows the fraction of the LHCb parameter space to the $p$He and He$p$ source terms. 
We make the conservative assumption that the cross section is only known in a small (roughly 10\%) range around the measured $\sS$. 
In this case, the contribution to the $p$He channel is at the permille level, peaking at an energy between between 10 and 100 GeV, 
while the contribution to the He$p$ channel is significantly larger at the 4\% level. 
The different coverage of the $p$He source spectrum in the inverse He$p$ kinematic configuration depends on the fact that  in the CM frame all but one 
LHCb data point correspond to backwards scattering in the $p$He system, or equivalently forwards scattering in the He$p$ system. 
The source term integral in \eqnref{sourceTerm_1} enhances the high-energy forward scattering due to the convolution with the steeply falling CR flux.
Since in any case the contribution of the LHCb data to the source terms is very small,  it is impossible to base 
the calculation of the $p + {\rm He} \rightarrow \bar{p} +X$ production solely on LHCb data. 
In the  parameterization of  the $p$He cross section,  we will therefore rely on a re-scaling of the  $pp$ ruled by  
the $p$C data from NA49 \cite{Baatar:2012fua}, taken at $\sS=17.3$~GeV. 
Their contribution to the source term, as visible in \figref{contribution_pHe},  is comparable in energy and amount to the $pp$ contribution from NA49.

The important conclusion from \figref{contribution_pHe} is that the current LHCb data are not yet sufficient to give a full picture of the 
the antiproton production spectrum in the helium channels and its uncertainties. 
The contribution of the incoming $p$ or He at the highest energy contribute only a small fraction to the 
produced antiprotons, in particular, referring to AMS-02 energies.
This result is due to the fact that during the computation of the source spectrum  the cross section is folded with an incident beam, namely the CR flux, which follows an energy power law with index of about $-2.7$. 
Nonetheless, the LHCb data contain valuable information: 
It shows for the first time how well the rescaling from the $pp$ channel applies to a helium target and how the cross section extrapolation to 
high energies works. 
Moreover, finding an agreement between LHCb data and predictions based on $pp$ and $p$C,
 increases trust in our current approaches and models. 
The way to improve the contribution of LHCb and the significance of its data is to increase the antiproton detection threshold above 100~GeV and/or lowering the incident proton energy below 1~TeV. 
In  \appref{app_source_term_fraction} we present predictions for the contribution with LHCb data at lower CM energies. Furthermore, 
we give an update of the results from DKD17  in \appref{app_par_space} to determine the  
whole relevant parameter space of $pA$ cross sections to interpret AMS-02 data. The update takes into account the asymmetry of the 
cross section, namely it is given in terms of $\xF$ instead of $\xR$.

\section{\label{sec::pp_fit} Fitting the proton-proton channel} 

The proton-proton channel is relevant since it contributes about 40\% of the total and, furthermore, it is 
the baseline for re-scaling to heavier nuclei, and for treating the contribution from antineuterons and hyperons. 
Its accurate determination is of central importance, since any uncertainty in $pp$ directly translates into all the other channels.
In the following we test and update the most recent analytic parametrizations 
by Di Mauro \etalS \cite{diMauro:2014_pbarCrossSectionParametrization} and 
Winkler \cite{Winkler:2017xor}, employing the  NA49 \cite{NA49_Anticic:2010_ppCollision} and 
the newly available NA61 data \cite{Aduszkiewicz:2017sei}. To reduce systematic biases we will try to discard most of the old data sets. 
Before turning to the fit results, we devote separate discussions to hyperons and isospin violation, 
the cross section parameterizations, the cross section data sets, and the fitting procedure. 

\subsection{Isospin violation and hyperons}
The fits that we are going to perform are on the prompt antiproton production, so that  antineutrons or antihyperons 
which subsequently decay into antiprotons are excluded from the fit. 
The estimate of the antiproton source term in the Galaxy requires the addition of these contributions by re-scaling from the prompt production
\begin{eqnarray}
  \label{eqn::XS_galaxy}
  \sigmaInv^\mathrm{Galaxy} = \sigmaInv (2 + \Delta_{\mathrm{IS}} + 2 \Delta_{\Lambda}),
\end{eqnarray}
where $\Delta_{\mathrm{IS}}$ is the enhancement factor of antineutron with respect to antiproton production 
 and $\Delta_\Lambda$  is the hyperon factor\footnote{We assume that the antiproton and antineutron production from hyperons is equal.}. 
The investigations in \cite{Winkler:2017xor} indicate that the factors $\Delta_{\mathrm{IS}}$ and $\Delta_\Lambda$ are energy dependent. 
We adopt these results and shortly repeat the analytic formulas for completeness: 
\begin{eqnarray}
  \label{eqn::isospin}
  \Delta_\mathrm{IS} &=& \frac{c_1^\mathrm{IS}}{1+(s/c_2^\mathrm{IS})^{c_3^\mathrm{IS}}},
\end{eqnarray}
with 
$c_1^\mathrm{IS} = 0.114$, $c_2^\mathrm{IS} = (144\,\mathrm{GeV})^2$, and $c_3^\mathrm{IS} = 0.51$
and 
\begin{eqnarray}
  \label{eqn::hyperon}
  \Delta_\Lambda &=& 0.81 \left(c_1^\Lambda + \frac{c_2^\Lambda}{1+(c_3^\Lambda /s)^{c_4^\Lambda}} \right),
\end{eqnarray}
with $c_1^\Lambda = 0.31$, $c_2^\Lambda = 0.30$, $c_3^\Lambda = (146\,\mathrm{GeV})^2$, and $c_4^\Lambda = 0.9$.
The uncertainties of these parameters have been determined in  \cite{Winkler:2017xor}. 
Their impact on the antiproton spectrum is discussed later in this paper. 

\subsection{Cross section parametrization}

We use two parameterizations in the fit: Eq.~(12) by Di Mauro  \etalS \cite{diMauro:2014_pbarCrossSectionParametrization} (hereafter Param. I)
and Winkler~\cite{Winkler:2017xor} (Param. II). 
Both formulae are given for the Lorentz invariant cross section in the CM frame as a function of 
the kinetic variables $\sS$, $\xR$, and $\pT$. 
Param.~I depends on 8 fit parameters $\mathcal{C}=\lbrace C_1...C_8 \rbrace$
\begin{eqnarray}
  \label{eqn::param_diMauro}
  \sigmaInv (\sS, \xR, \pT) &=&\sigma_{\mathrm{in}} (1-\xR)^{C_1} \exp(-C_2 \xR)           \\ \nonumber 
              &&\times \left[ C_3\left(\sS\right)^{C4} \exp(-C_5 \pT) \right.  \\ \nonumber
            && \left.+ C_6\left(\sS\right)^{C7} \exp\left(-C_8 \pT^2\right) \right].
\end{eqnarray}
The pre-factor $\sigma_{\mathrm{in}}$ is the total inelastic $pp$ cross section and its energy-dependent form is given 
in~\cite{diMauro:2014_pbarCrossSectionParametrization} (Appendix~B). 
We note that  this parametrization allows freedom for the scaling with $\sS$ and $\pT$.  
Especially, it includes an increasing normalization $\sigma_{\mathrm{in}}(s)$ which is determined by a separate fit to data. 

Param.~II depends only on 6 parameters $\mathcal{C}=\lbrace C_1...C_6\rbrace$ and is given by
\begin{eqnarray}
  \label{eqn::param_Winkler}
  \sigmaInv (\sS, \xR, \pT) &=& \sigma_{\mathrm{in}} R \, C_1  (1-\xR)^{C_2} \\ \nonumber
  &&\times \left[ 1 + \frac{X}{\mathrm{GeV}}(m_T - m_p) \right]^{\frac{-1}{C_3 X}},
\end{eqnarray}
where $m_T=\sqrt{p_T^2 + m_p^2}$. 
The factor 
\begin{eqnarray}
  \label{eqn::param_Winkler_III}
   R  &=& 
   \begin{cases} 
     1 & \sS \geq 10\,\mathrm{GeV} \\
     \left[ 1 + C_5 \left(10-\frac{\sS}{\mathrm{GeV}}\right)^5 \right]     &  \text{elsewhere}        \\ 
     \quad \times \exp \left[ C_{6} \left( 10 - \frac{\sS}{\mathrm{GeV}} \right)^2 \right. \\ 
     \quad \times \left. (\xR - x_{R,\mathrm{min}} )^2  \right]
   \end{cases}
\end{eqnarray}
describes the scaling violation of the cross section at low $\sS$, and $x_{R,\mathrm{min}}=m_p/E_\pb^{\mathrm{max}*}$.
As before, $\sigma_{\mathrm{in}}$ is the total inelastic cross section, whose form is determined to be
\begin{eqnarray}
  \label{eqn::param_Winkler_II}
  \sigma_{\mathrm{in}} &=& c_{\mathrm{in},1} + c_{\mathrm{in},2} \log\left( \sS \right) + c_{\mathrm{in},3} \log^2\left( \sS \right), \end{eqnarray}
with $c_{\mathrm{in},1} = 30.9$~mb, $c_{\mathrm{in},2} = -1.74$~mb, and $c_{\mathrm{in},3} = 0.71$~mb.
Finally, the last factor of \eqnref{param_Winkler} describes the scaling violations at large $\sS$. This factor contains the parameter
\begin{eqnarray}
  \label{eqn::param_Winkler_IV}
  X &=& C_4 \log^2 \left( \frac{\sS}{4 m_p} \right).
\end{eqnarray}
The scaling violation at large energies affects the cross section parametrization in two ways. 
Firstly, the total inelastic $pp$ cross section rises according to \eqnref{param_Winkler_II} and, secondly, 
the $\pT$ shape is changed as described by the last factor of \eqnref{param_Winkler}. 
Scaling violations were intensively studied in by Winkler~\cite{Winkler:2017xor} and found not to affect the behavior 
of the cross section below $\sS=50$~GeV. 
In this analysis we are interested in low-energy part, where NA61 adds new data. A closer look at \eqnref{param_Winkler} reveals that the parameter $C_3$ determines the $\pT$ shape at low energies, 
while $C_4$ regulates the strength of alteration towards high energies. 
So, we  fix the parameter $C_4=0.038$ \cite{Winkler:2017xor}, while allowing the other 5 parameters to vary freely.  

\subsection{Data}

The main data sets to constrain the fit on $\sigma_{p+p\rightarrow \bar{p}+X}$ are the NA49  \cite{NA49_Anticic:2010_ppCollision} and 
 NA61 \cite{Aduszkiewicz:2017sei} ones. 
However, the discussion about \figref{contribution_pp}  revealed the necessity of a further
data set at low energies to fix the antiproton source term below $\Tpbar = 5$~GeV. 
There are only two available data sets at these energies: Dekkers \etalS \cite{Dekkers:1965zz} taken at $\sS=6.1$ and 6.7~GeV and
Allaby \etalS~\cite{Allaby:1970jt}  at $\sS=6.15$~GeV.
We use  the measurements by Dekkers \etalS, while
the data set by Allaby \etalS~\cite{Allaby:1970jt} is not 
taken into account because it contains very small statistical errors in combination with large systematic and normalization uncertainties. 
When fitting Param.~I, we add data from the BRAHMS experiment, which is taken in $pp$ collisions at $\sS=200$~GeV \cite{Arsene:2007jd}, 
in order to fix the freedom of the high-energy behavior in this parameterization. 
In the case of Param.~II, we fixed the high-energy behavior (see discussion above) and, thus, 
the additional data set is not necessary. A summary of all $pp$ data is given in \tabref{pp_data}.
\begin{table}
\caption{ Summary of all $pp$ data sets, their available CM energies, and references. 
          Moreover, we declare which parametirzation (I~or~II) is used and which scale uncertainty
          $\sigma_{\mathrm{scale}}$ is adopted in the fits (see \eqnref{chiSquare_scale}). }
\label{tab::pp_data}
\begin{tabular}{l c c c c c}
  \\ \hline \hline
  Experiment & $\sS\quad \mathrm{[GeV]}$ & $\sigma_{\mathrm{scale}}$ & I & II & Ref.  \\ \hline
  NA49 & 17.3 & 6.5\% & $\times$ & $\times$ & \cite{NA49_Anticic:2010_ppCollision} \\
  NA61 & 7.7, 8.8, 12.3, 17.3 & 5\% & $\times$ & $\times$ & \cite{Aduszkiewicz:2017sei} \\
  Dekkers \etal  & 6.1, 6.7 & 10\% & $\times$ & $\times$ & \cite{Dekkers:1965zz} \\
  BRAHMS & 200  & 10\% & $\times$ &  & \cite{Arsene:2007jd} \\ \hline \hline
\end{tabular}
\end{table}
The NA49 and NA61 collaborations  explicitly determine the prompt antiproton flux, namely, hyperon-induced antiprotons are subtracted
from the original data. 
However, for older experiments the situation is not completely clear. Since hyperons have a very short life-time, 
they usually decay inside the detector and can contribute to the measurement. Modern detectors, such as NA49, NA61 and LHCb, 
can reconstruct a primary vertex and discard hyperon-induced antiprotons. 
The usual assumption for older experiments is that they did not distinguish between hyperon-induced and prompt antiprotons. 
Thus, to use their data, in our case Dekkers \etalS and BRAHMS, we subtract the hyperon contribution according to \eqnref{hyperon}.
Since antineutrons have a far longer lifetime compared to hyperons, they never decay inside the detector and do not require a similar correction.

\subsection{Fit Procedure} 

We perform a $\chi^2$-fit using the \textsc{Minuit} package from \textsc{Root}\footnote{\url{https://root.cern.ch}} software framework in order to minimize the $\chi^2$, 
which is divided into two terms:
\begin{eqnarray}
  \label{eqn::chiSquare}
  \chi^2 (\mathcal{C}, \omega )               &=& \chi_\mathrm{stat}^2 (\mathcal{C}, \mathcal{\omega})   +   \chi_\mathrm{scale}^2(\mathcal{\omega}).
\end{eqnarray}
The first term accounts the statistical information and contains a sum over all the data points $i_k$ of all the experiments $k$ from 1 to $L$:
\begin{eqnarray}
  \label{eqn::chiSquare_stat}
  \chi_\mathrm{stat}^2 (\mathcal{C}, \omega) &=& \sum \limits_{k=1}^{L} \sum\limits_{i_k}                         
                                         \frac{\left( \omega_k {\sigmaInv}_{,i_k} - 
                                         \sigmaInv(  \mathcal{C},  \mathcal{T} )_{,i_k}     \right)^2}{\omega_k^2 \sigma_{i_k}^2}. \quad
\end{eqnarray}
Here ${\sigmaInv}_{,i_k}$ is the $i_k$  data point  for invariant cross section having total uncertainty $\sigma_{i_k}$, which is taken as 
 the quadratic sum of  statistical and systematical uncertainties of each data point if both are stated separately.  
The cross section parametrization $\sigmaInv( \mathcal{C},\mathcal{T} )$ is evaluated (either for Param.~I or II) at the parameter set $\mathcal{C}$ 
and the kinematic variables 
$ \mathcal{T} = {\sS}^{(i_k)}, {\xR}^{(i_k)}, {\pT}^{(i_k)}$.
For each data set we allow a re-scaling  by a constant factor $\omega_k$, which penalizes the $\chi^2$
 by the second term in \eqnref{chiSquare}, explicitly  given by 
\begin{eqnarray}
  \label{eqn::chiSquare_scale}                                       
  \chi_\mathrm{scale}^2(\omega ) &=& \sum \limits_{k=1}^{L}
                                           \frac{\left( \omega_k - 1 \right)^2}{ \sigma_{\mathrm{scale},k}^2},
\end{eqnarray}
where $\sigma_{\mathrm{scale},k}$ is the scale uncertainty for each data set (see \tabref{pp_data}). Note that the sum in \eqnref{chiSquare_stat} runs over every single data point, while the sum in \eqnref{chiSquare_scale} only runs over all data sets. Indeed, a scale uncertainty requires that all points are correlated. Moving up or down all the data points of each set  by the same factor is only penalized once, not for each data point. This is in contrast to the treatment in \cite{diMauro:2014_pbarCrossSectionParametrization} (although stated differently in the paper).  NA49  explicitly states scale uncertainties of 6.5\%.
For the other experiments we estimate the scale uncertainty according to the average size of the systematic uncertainties to be 
5\% for NA61, and 10\% for Dekkers \etalS and BRAHMS. 
During the fit we simultaneously adjust $\mathcal{C}$ and the nuisance parameters $\omega=\omega_1 ... \omega_L$, leaving in total 12 free parameters for Param.~I and 8 for Param.~II. In practice, we use the \textsc{Root} algorithms in two steps to minimize our $\chi^2$. First, the MIGRAD algorithm determines a good parameter estimate, then the HESSE algorithm confirms  these parameters and gives a more reliable covariance matrix.

\subsection{Results}

\begin{table}[b!]
\caption{  Fit results to the $p+p\rightarrow\pbar + X $ cross section. 
           The full correlation matrices are given in \tabref{Correlation_pp_diMauro} and \tabref{Correlation_pp_Winkler}. \\
           $^{(*)}$ The parameter $C_4$ is fixed, {\it i.e.} not included in the fit, in Param.~II (see text for details). }
\label{tab::Fit_results_pp}
\begin{tabular}{ l  c  c }
 \hline \hline
                         &   with Param.~I                   &   with Param.~II                   \\ \hline
$C_1$                     &   $3.50\pm 0.64$                  & $(5.02\pm 0.22)\times 10^{-2}$     \\
$C_2$                     &   $5.59\pm 0.85$                  & $7.790\pm0.077$                    \\
$C_3$                     &   $(4.00\pm 0.73)\times 10^{-2}$  & $(1.649\pm 0.012)\times 10^{-1}$   \\
$C_4$                     &   $-0.251\pm0.071$                & ${\it (3.800\pm 0.057)\times 10^{-2} \,}^{(*)}$   \\
$C_5$                     &   $2.651\pm0.097$                 & $(4.74\pm 2.59)\times 10^{-4}$     \\
$C_6$                     &   $(3.78\pm 0.53)\times 10^{-2}$  & $3.70\pm0.64$                      \\
$C_7$                     &   $(4.3\pm 4.3)\times 10^{-2}$    & - \\
$C_8$                     &   $2.695\pm0.047$                 & - \\
$\omega_\mathrm{BRAHMS}$  &   $1.115\pm0.079$                 & - \\
$\omega_\mathrm{Dekkers}$ &   $1.051\pm0.068$                 & $1.090\pm0.090$                 \\
$\omega_\mathrm{NA49}$    &   $1.059\pm0.039$                 & $1.061\pm0.044$                 \\
 $\omega_\mathrm{NA61}$   &   $0.936\pm0.036$                 & $0.932\pm0.038$                 \\  \hline \hline
\end{tabular}
\end{table}

We find that both parametrizations \eqnref{param_diMauro} and~\eqref{eqn::param_Winkler}, result in a good fit to the $pp$ cross section data, giving a 
$\chi^2$/ndf of 1.30 and 1.18 for Param.~I and Param.~II, respectively. The best fit parameters with the relevant 
$1\sigma$ uncertainties are summarized in~\tabref{Fit_results_pp}. We present the full correlation matrix in the Appendix. 
Furthermore, we demonstrate in \tabref{Fit_properties_pp} that 
reduced $\chi^2$s for all data sets are close to one, namely, 
all data sets in the fit are consistent with
each other, if we allow for a re-scaling by the nuisance parameters $\omega_k$.
There seems to be a general tendency to scale down the NA61 data by 6-7\%, while increasing the NA49 data by about 6\% and
the Dekkers data by 5-9\% in both parametrizations. 
 Especially, we note that NA49 and NA61 have an overlap in the kinetic parameter space at $\sS=17.3$~GeV. After the slight re-scaling the two data sets are in agreement with each other.

\begin{table}[t!]
\caption{Fit quality of the $pp$ channel. The first row reports the global fit, while the other ones show the contribution of the single data sets to the $\chi^2$. }
\label{tab::Fit_properties_pp}
\begin{tabular}{ l c c }
\hline \hline
                                          &   with Param.~I  &   with Param.~II   \\ \hline
$\chi^2$/ndf                            &   534.7/411      &   464.7/394    \\
$\chi^2_\mathrm{BRAHMS}$ (data points)    &   27.6   (21)    &       -        \\
$\chi^2_\mathrm{Dekkers}$(data points)    &    9.8   (10)    &   8.3    (10)  \\
$\chi^2_\mathrm{NA49}$   (data points)    &   211.4 (143)    &   179.0 (143)  \\
$\chi^2_\mathrm{NA61}$   (data points)    &   286.0 (249)    &   277.4 (249)  \\  \hline \hline
\end{tabular}
\end{table}

\begin{figure}[b!]
	\includegraphics[width=1.\linewidth]{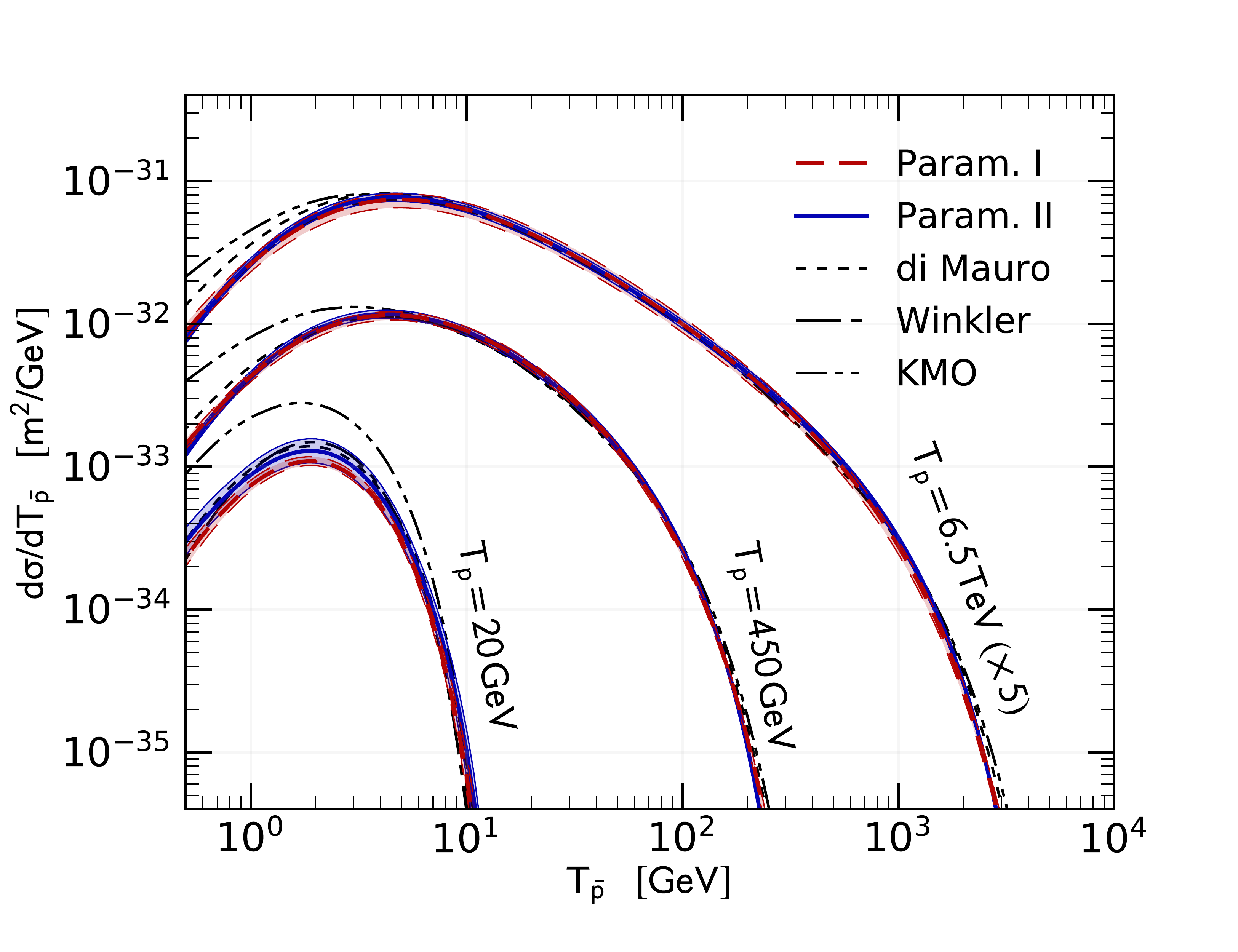}
  \caption{ The differential cross section $d\sigma/dT_{\pbar} (p+p\rightarrow \pbar + X)$ for prompt antiprotons, 
  			 at the representative proton energies $T_p$ = 20 GeV, 450 GeV and 6.5 TeV. 
  			 The dashed (solid) line and the red (blue) band are the result of our analysis for Param. I and Param. II.
  			 The uncertainty band corresponds to the $2\sigma$ confidence interval.
  	         We report for comparison some literature estimations (see text for details). 
  			 Tables with the full cross section results are provided in the Supplemental Material to this paper. } 
\label{fig::comparison_XS_pp}
\end{figure}

\begin{figure}[b!]
	\includegraphics[width=1.\linewidth, trim={0  4.8cm 0 0},clip]{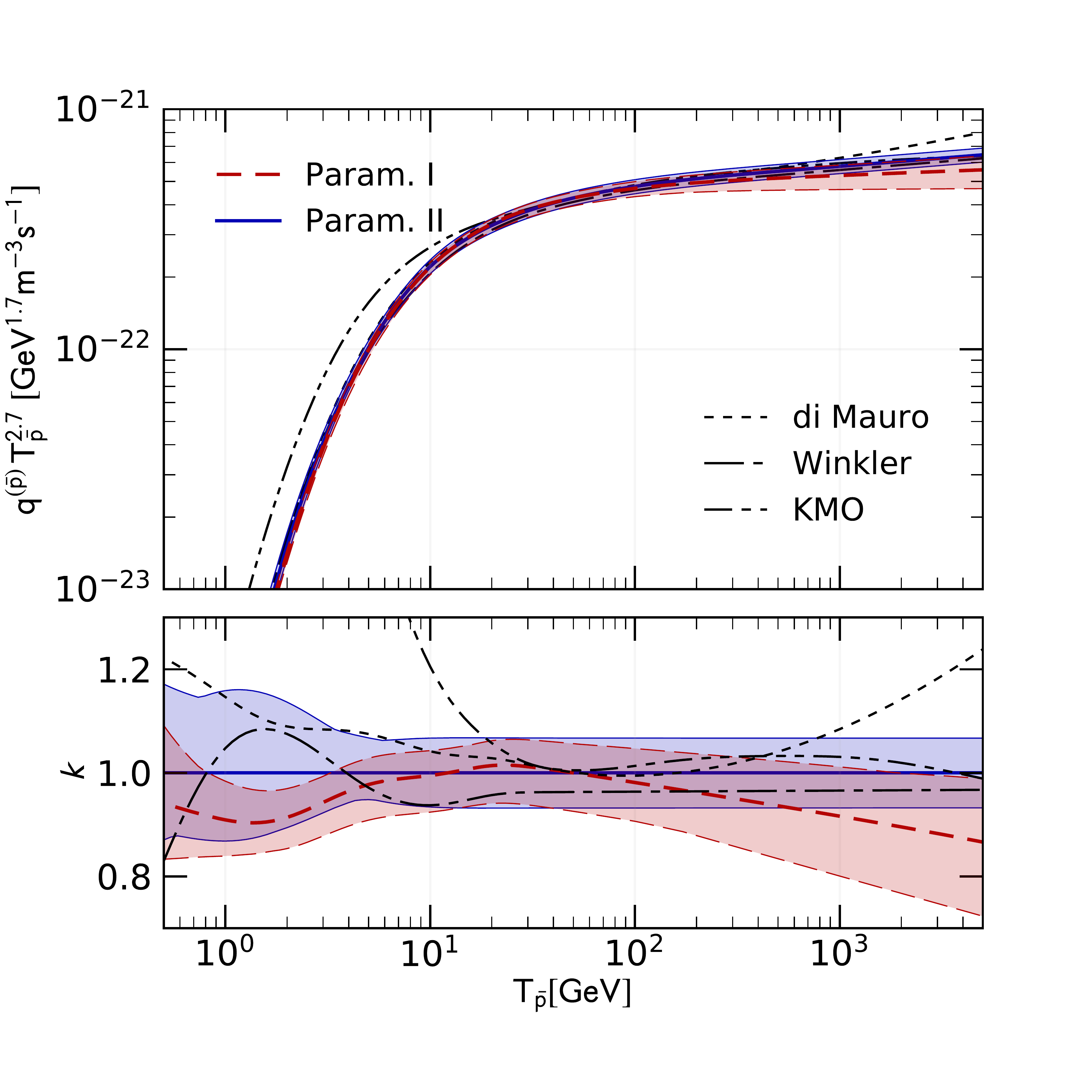}
	\includegraphics[width=1.\linewidth, trim={0 0 0 18.2cm},clip]{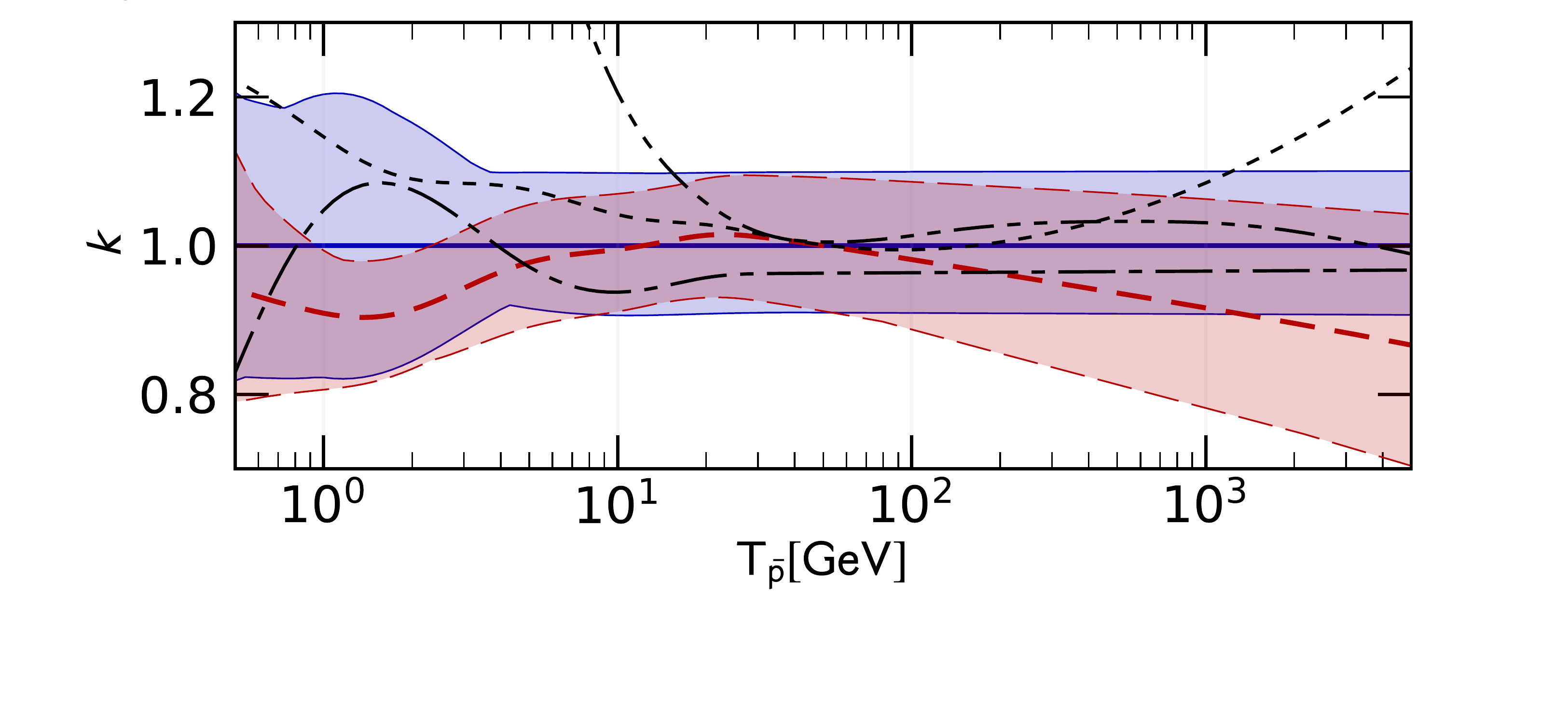}
  \caption{ Source term of prompt antiprotons originating from $pp$ collisions and its uncertainty induced by 
            the cross section fits of Param.~I (red dashed) and Param.~II (blue solid), respectively. 
            For comparison, we show the antiproton source term from previous parametrizations Di Mauro \etal , Winkler and KMO. 
            The central panel displays the ratio $k$ to the best fit of Param.~II
            and the shaded uncertainty bands correspond to the $2\sigma$ confidence interval. 
            For completeness the lower panel contains the $1\sigma$ envelope of the $n$-dimensional $\chi^2$ distribution
            (see discussion in the text for details). 
          }
	\label{fig::comparison_pp}
\end{figure}

In \figref{comparison_XS_pp} we report our results for the differential cross section 
$d\sigma/dT_{\pbar} (p+p\rightarrow \pbar + X)$ for the production of prompt antiprotons, at the representative proton energies   $T_p = 20$ GeV, 450 GeV and 6.5 TeV. 
The uncertainty bands are derived from our fits (see discussion about \figref{comparison_pp}), and go from 20\% at $T_p = 20$ GeV to 10\% for  $T_p =450$ GeV and 6.5 TeV. 
For the lower $T_p$ value the to predictions are mildly compatible, while they 
almost overlap for higher energies. We also report some estimates from recent literature, showing some discrepancy with our 
findings mostly at low $T_{\pbar}$ for the Di Mauro and the Winkler parameterizations. The Monte Carlo based KMO parameterization has been divided by 2.3 since 
it accounts all the antiprotons produced in the interaction, {\it i.e.} not only the prompt ones. 
We provide tables with the total (antiprotons from prompt production, from antineutron and antihyperon dacay) cross section for a full $T_p$ scan - as well as for a number of other incoming particles on $p$ and He -  in the Supplemental Material.

We calculate the $\bar{p}$ source term from the two cross section parameterizations, and compare them two previous predictions.
We remind again that the fit is performed to the prompt antiproton production, and consequently the source term
calculated according to \eqnref{sourceTerm_1} and displayed in \figref{comparison_pp} does not include antiprotons from neutron and hyperon decay.
To calculate the  fit uncertainty we sample random points in the parameter space $(\mathcal{C},\omega)$ from the full correlation matrix and verify each point against the total $\chi^2$ from \eqnref{chiSquare}.
Then we compute the $\chi^2$ profile as function of the source term, separately at each energy. The uncertainty band at 1$\sigma$ ($n\sigma$) is given by $\Delta \chi^2 = 1$ ($\Delta \chi^2 = n^2$). The interpretation of this $1\sigma$ region is that in 68.3\% of all cases the source term falls within the band.  We checked that the size of the uncertainty band grows approximately linearly with the $\sigma$-interval. Therefore, we show only the $2\sigma$ band in our plots, a different confidence level may be obtained by rescaling.
An alternative useful quantity is the envelope of the $n$-dimensional $\chi^2$ distribution at $1\sigma$ level, where $n$ is the number of free parameters in the fit (Param.~I: $n=12$, Param.~II: $n=8$). In other words, this envelope is built from the set which contains of 68.3\% of the source term realizations and has the lowest $\chi^2$ values. We show this envelope in \figref{comparison_pp} for comparison and note that it almost coincides with the $3\sigma$ band.   
Finally, we obtain an uncertainty - solely from the cross section fit - of $\pm8$\%. With Param.~II it increases to about $\pm15$\% below 5~GeV. 
The source terms from Param.~I and Param.~II are consistent within the fit uncertainties.
In particular, from $\Tpbar=1$~GeV to a few hundred GeV, the agreement between the two models
is very good. Above 500~GeV, Param.~II provides an antiproton spectrum systematically higher than Param.~I. 
In~\cite{diMauro:2014_pbarCrossSectionParametrization}, it was already pointed out that Param.~I - due to the employed data sets - 
gives  reliable results  up to a few hundred GeV. 
Param.~II, which employs different data sets at the highest $\sS$,  is likely to give a more trustworthy result at high energies. 
The comparison of previous analysis by Di~Mauro, Winkler and KMO reveals several interesting features. 
The direct (because involving the same parametrization, just different data sets) comparison between Param.~I and Di Mauro shows that the source term predictions
are very close between 10 and 100~GeV, while Param.~I source spectrum stands systematically lower below 10~GeV and above 100~GeV. 
This is probably the effect of the hyperon subtraction for Dekkers and BRAHMS data, which was considered here but not 
in~\cite{diMauro:2014_pbarCrossSectionParametrization}. Param.~II and Winkler are consistent within uncertainties. Especially, above 30~GeV
the two predictions also coincide with KMO. Below 10~GeV KMO clearly deviates and overpredicts the antiproton source spectrum.

Concluding, both parametrizations for the $\sigma_{p+p\rightarrow \pbar + X}$  result in compatible $\pbar$ source terms for the energy range measured by AMS-02. The difference in the shape 
of the two parametrizations is only few percent in the range of $\Tpbar= 5$ to 100~GeV, however, at 1~GeV and 1~TeV it grows to 10\%, where Param.~I is slightly softer at high energies.

\section{\label{sec::pA_fit} Fitting the proton-nucleus channel}

The proton-nucleus channels contribute between 40 and 50\% of the total secondary antiproton production.
However, the currently available data on  antiproton production measurement in $pA$ collisions are not sufficient to 
allow an individual description of each relevant channel, especially $p$He (see discussion in \secref{current_status}).
We use $p$C data by NA49 and $p$He data by LHCb to determine a re-scaling factor for the $pA$ and, specifically, 
$p$He cross sections from the $pp$ cross section.

\subsection{Cross section parametrization} 
Antiproton production in $pp$ collisions is by definition symmetric under a reflection along the beam axis in the CM frame, 
while this is not necessarily the case in $pA$ collisions (in the nucleon-nucleon CM frame). 
Actually, NA49 $p$C data~\cite{Baatar:2012fua} reveals that the cross section is not symmetric between forward and backward production. 
It is plausible that the binding of the nucleons in the nucleus has an effect on the antiproton production and breaks the symmetry. 
Since a description of the cross section in terms of $\xR$ which intrinsically expects symmetry is inconsistent, we will use $\xF$ instead in the following whenever we discuss $pA$ channels  .
Following the description by NA49, \cite{Kappl:2014_pbarCrossSection} exploits a re-scaling of $pp$ cross section in terms of overlap functions. 
The idea is to split the antiproton production into two components produced by projectile and target,  
where the antiprotons from each component are produced mainly forward directed. Separately adjusting the overlap functions allows to accommodate the asymmetry.
\begin{figure*}[t!]
	\includegraphics[width=.5\linewidth]{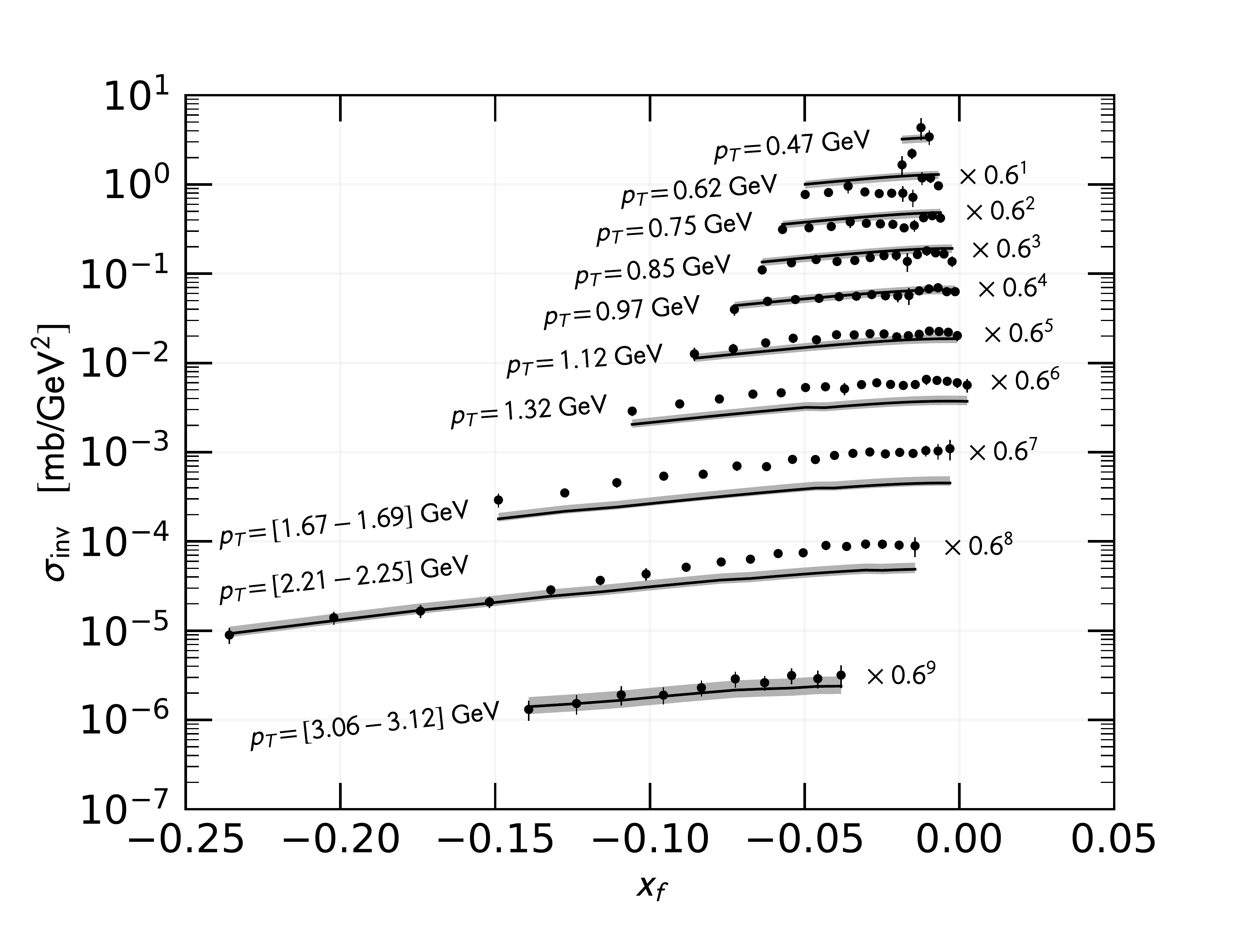}\includegraphics[width=.5\linewidth]{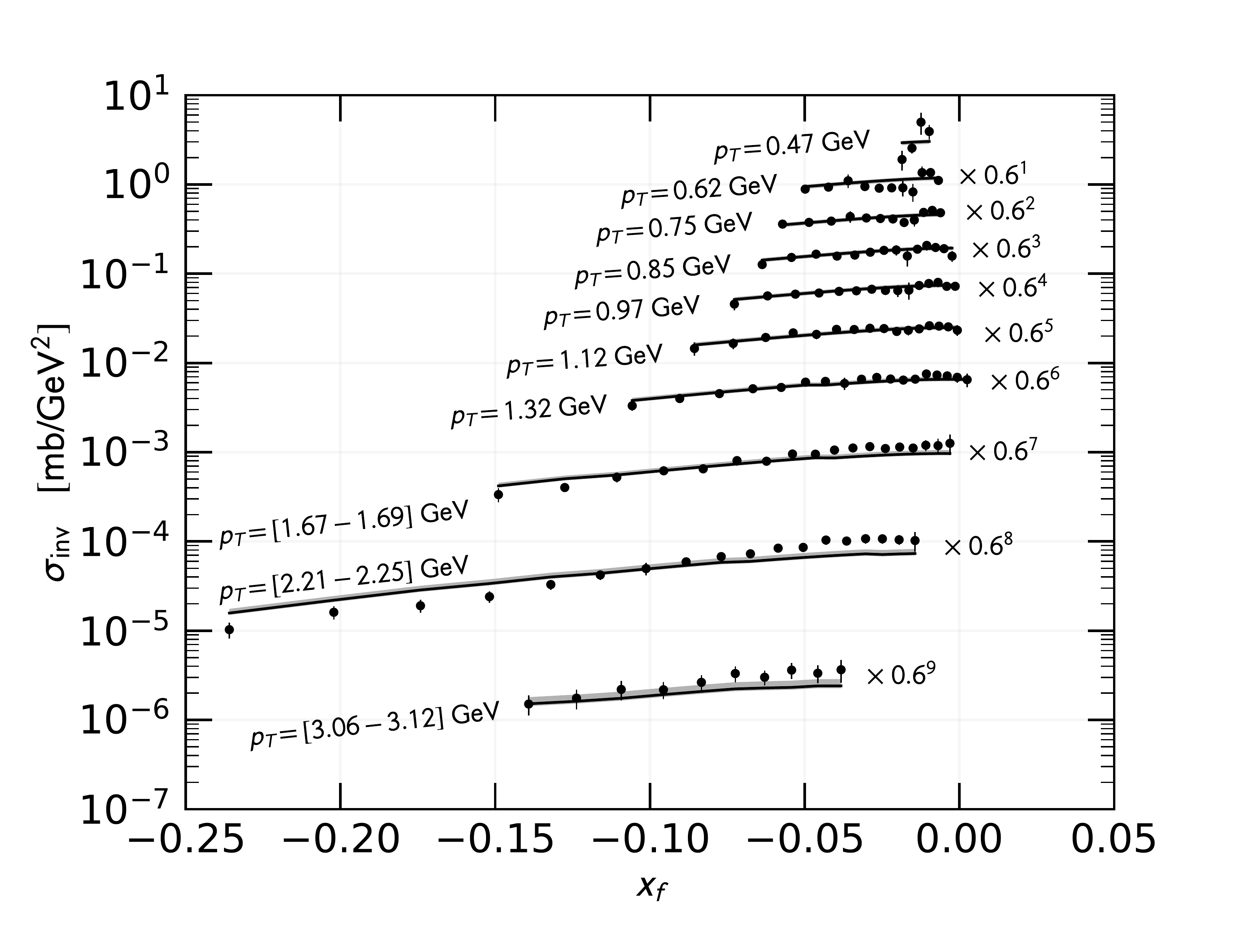}
  \caption{ Comparison of LHCb data to the fit with Param.~I-B (left) and Param.~II-B (right). 
            The grey band corresponds to 2$\sigma$ uncertainty in the fit. The LHCb data agree better with Param.~II and, therefore,  they 
            select this model for the high-energy behavior of the Lorentz invariant cross section.
          }
	\label{fig::LHCb_data}
\end{figure*}
\begin{table}[b!]
\caption{Projectile overlap function $F_\mathrm{pro}(\xF)$. The definition is taken from \cite{Baatar:2012fua}. }
\label{tab::Overlapfunction}
\begin{tabular}{ c l | c l}
\hline \hline
 $\xF$     &  $F_\mathrm{pro}$&  $\xF$     &  $F_\mathrm{pro}$   \\ \hline
-0.250     &   0.0000  &  0.250     &   1.0000  \\ 
-0.225     &   0.0003  &  0.225     &   0.9997  \\ 
-0.200     &   0.0008  &  0.200     &   0.9992  \\ 
-0.175     &   0.0027  &  0.175     &   0.9973  \\ 
-0.150     &   0.010   &  0.150     &   0.990   \\ 
-0.125     &   0.035   &  0.125     &   0.965   \\ 
-0.100     &   0.110   &  0.100     &   0.890   \\ 
-0.075     &   0.197   &  0.075     &   0.803   \\ 
-0.050     &   0.295   &  0.050     &   0.705   \\ 
-0.025     &   0.4     &  0.025     &   0.6     \\ 
           &           &  0.000     &   0.5     \\ \hline \hline
\end{tabular}
\end{table}
The inclusive Lorentz invariant cross section of $p + A \rightarrow \pbar + X$ scattering is given by
\begin{eqnarray}
  \label{eqn::param_pA}
  \sigmaInv^{pA} (\sS, \xF, \pT) &=& f^{pA}(A, \xF, \mathcal{D}) \,\,  \sigmaInv^{pp}(\sS, \xR, \pT), \quad
\end{eqnarray}
where $A$ is the mass number of the nucleus and $\mathcal{D}=(D_1,D_2)$ are the two fit parameters. 
Explicitly, the factor $f^{pA}$ is defined by:
\begin{eqnarray}
  \label{eqn::param_pA_II}
  f^{pA} &=&  A^{D_1} \left[ A^{D_2} \left(1+\frac{N}{A}\Delta_\mathrm{IS} \right) F_\mathrm{pro}(\xF) \right. \\ \nonumber
         && \left. \qquad  \qquad +   F_\mathrm{tar}(\xF) \vphantom{\frac{N}{A}}  \right].
\end{eqnarray}
$F_\mathrm{pro}(\xF)$ and $F_\mathrm{tar}(\xF)$ are the projectile and target overlap functions.
They fulfil $F_\mathrm{tar}(\xF) = F_\mathrm{pro}(-\xF)$ and 
$F_\mathrm{tar}(\xF) + F_\mathrm{pro}(\xF) = 1$ and are defined in \tabref{Overlapfunction}. $N$ is the number of neutrons in the nucleus.
The form factor $f^{pA}$ is motivated by~\cite{Baatar:2012fua,Kappl:2014_pbarCrossSection}. Its $A$ dependence is chosen such that in the case of 
$A=1$ we retain proton-proton scattering. 
We remind that the kinetic variables $\xF$ and $\sS$ refer to the nucleon-nucleon CM frame, 
where proton and nucleus have the same velocity, not the same momentum. 
Consequently, the CM energy $\sS$ is the colliding energy of the nucleon with the proton.  

The fit procedure is analogous to the $pp$ case discussed in the previous section.
However, here the parameters $\mathcal{C}$ from \eqnref{param_pA} are fixed, in other words the $pp$ scattering is unaltered, 
and only the new parameters $\mathcal{D}$ are varied in the fit. 
The definition of our $\chi^2$ is equivalent to \eqnref{chiSquare}, when $\mathcal{C}$ is replaced by $\mathcal{D}$ and $k$ runs over the experiments with $pA$ data. As before we allow for nuisance parameters $\omega$ of each data set.

\subsection{Data}

\begin{table}[t!]
\caption{ Main properties of the NA49 $p$C and LHCb $pHe$ data sets: available CM energies, 
scale uncertainty $\sigma_{\mathrm{scale}}$ adopted in the fits, the parameterization (I~or~II) used in the fit, 
 references. Labels A and B refer to use of NA49 $p$C data alone or NA49 $p$C and LHCb $pHe$, respectively. }
\label{tab::pA_data}
\begin{tabular}{l c c c c c c c }
  \\ \hline \hline
       & $\sS\quad \mathrm{[GeV]}$ & $\sigma_{\mathrm{scale}}$ & I-A & I-B & II-A & II-B & Ref.  \\ \hline
  NA49 & 17.3 & 6.5\% & $\times$   & $\times$ & $\times$ & $\times$ & \cite{Baatar:2012fua} \\
  LHCb & 110  & 6.0\% &            & $\times$ &          & $\times$ & \cite{Graziani:2017} \\ \hline \hline
\end{tabular}
\end{table}

We exploit the data from NA49 and LHCb on $p$C and $p$He scattering, respectively.
Both experiments use a fixed target nucleus  while the incident proton is accelerated to 
158~GeV in NA49 and 6.5~TeV in LHCb. 
Their data are very precise, and both experiments correct for antihyperons, \textit{i.e.} they remove antiprotons originating from the decay of intermediate antihyperon states. 
We summarize the main experimental information in~\tabref{pA_data}.

\subsection{Results} 

\begin{table}[b!]
\caption{Fit quality of $f^{pA}$ for the different $pp$ Param.~I and II, and for the different data sets A (NA49 $p$C) and B (NA49 $p$C, LHCb $p$He). 
The first row shows the result of the fit, while the second and third rows report the split contribution from the $p$C NA49 and $p$He LHCb data sets. 
In brackets are the numbers of data points entering in the fit. The italic numbers are not the result of a minimization, but the  $\chi^2$ on LHCb data with the parameters fixed by NA49 $p$C data.}
\label{tab::Fit_properties_pA}
\begin{tabular}{ l c c  c c}
\hline \hline
                             &   \multicolumn{2}{c}{Param.~I}        &    \multicolumn{2}{c}{Param.~II}      \\
                             &    A                &    B            &    A                &    B            \\ \hline
$\chi^2$/ndf               &   153.0/118         &   1296.3/253      &   131.2/118         &   326.3/253     \\
$\chi^2_\mathrm{NA49}$       &   153.0 (121)       &   155.3 (121)   &   131.2 (121)       &   131.8 (121)   \\
$\chi^2_\mathrm{LHCb}$       & \textit{1266 (136)} &   1141 (136)    & \textit{212.4 (136)}&   194.5 (136)   \\ \hline \hline
\end{tabular}
\end{table}

\begin{table*}[t!]
\caption{  Fit results of $f^{pA}$ for the different $pp$ Param.~I and II, and for different data sets A (NA49 $p$C) and B (NA49 $p$C, LHCb $p$He).
           }
\label{tab::Fit_results_pA}
\begin{tabular}{ l c c c c }
\hline \hline
                          &   \multicolumn{2}{c}{Param.~I}            &    \multicolumn{2}{c}{Param.~II}              \\
Parameter                 &    A                  &    B              &    A                  &    B                  \\ \hline \hline
$D_1$                     &   $0.830\pm 0.012$    & $0.825\pm 0.012$  &   $0.825\pm 0.012$    & $0.828\pm 0.012$      \\
$D_2$                     &   $0.149\pm 0.013$    & $0.167\pm 0.012$  &   $0.154\pm 0.013$    & $0.145\pm 0.012$      \\
$\omega_\mathrm{NA49}$    &   $1.000\pm 0.025$    & $1.001\pm0.024$   &   $1.000\pm 0.025$    & $0.997\pm0.024$       \\
$\omega_\mathrm{LHCb}$    &   -                   & $0.900\pm0.015$   &   -                   & $1.034\pm0.018$       \\ \hline \hline
\end{tabular}
\end{table*}

\begin{figure*}[t!]
	\includegraphics[width=0.5\textwidth]{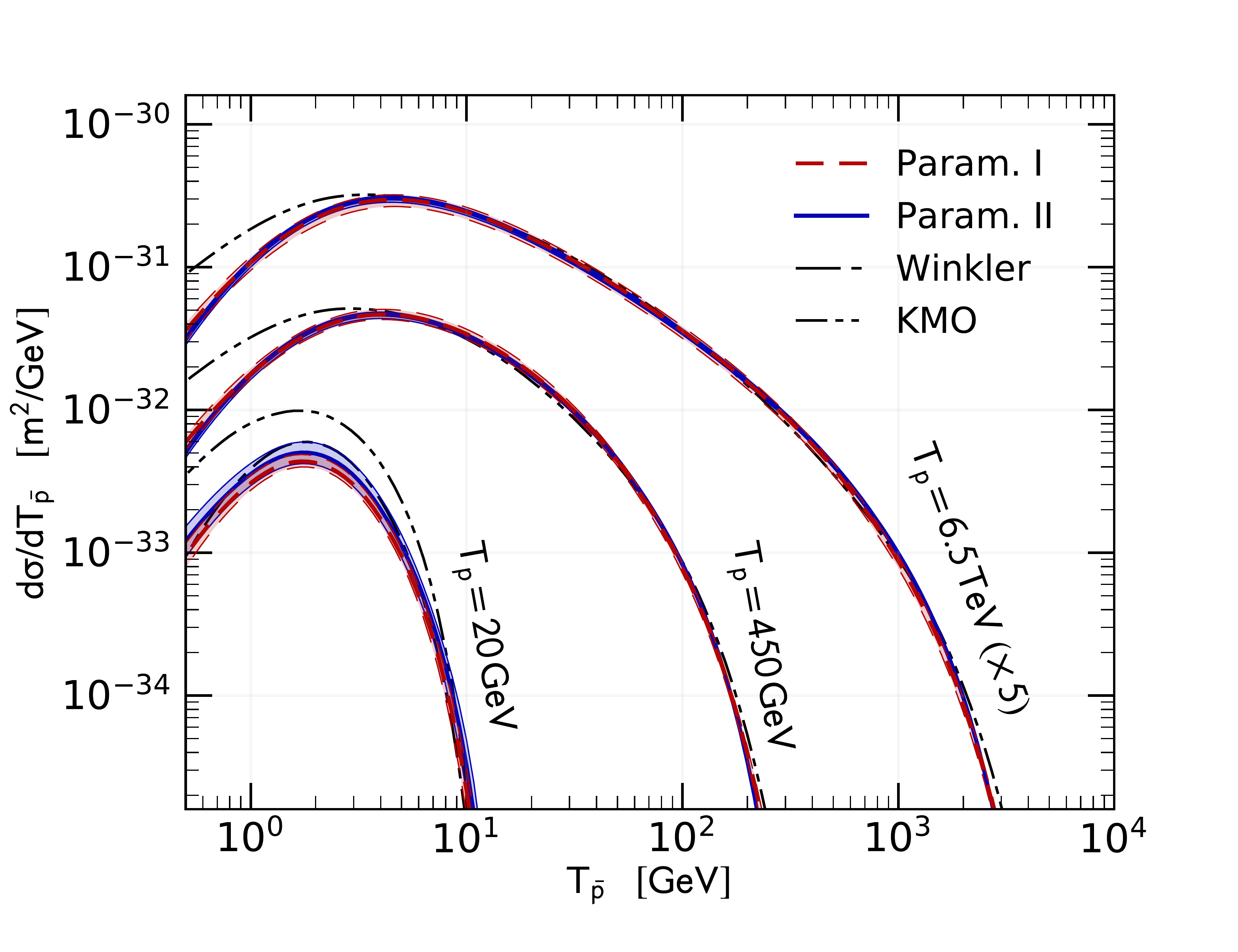}\includegraphics[width=0.5\textwidth]{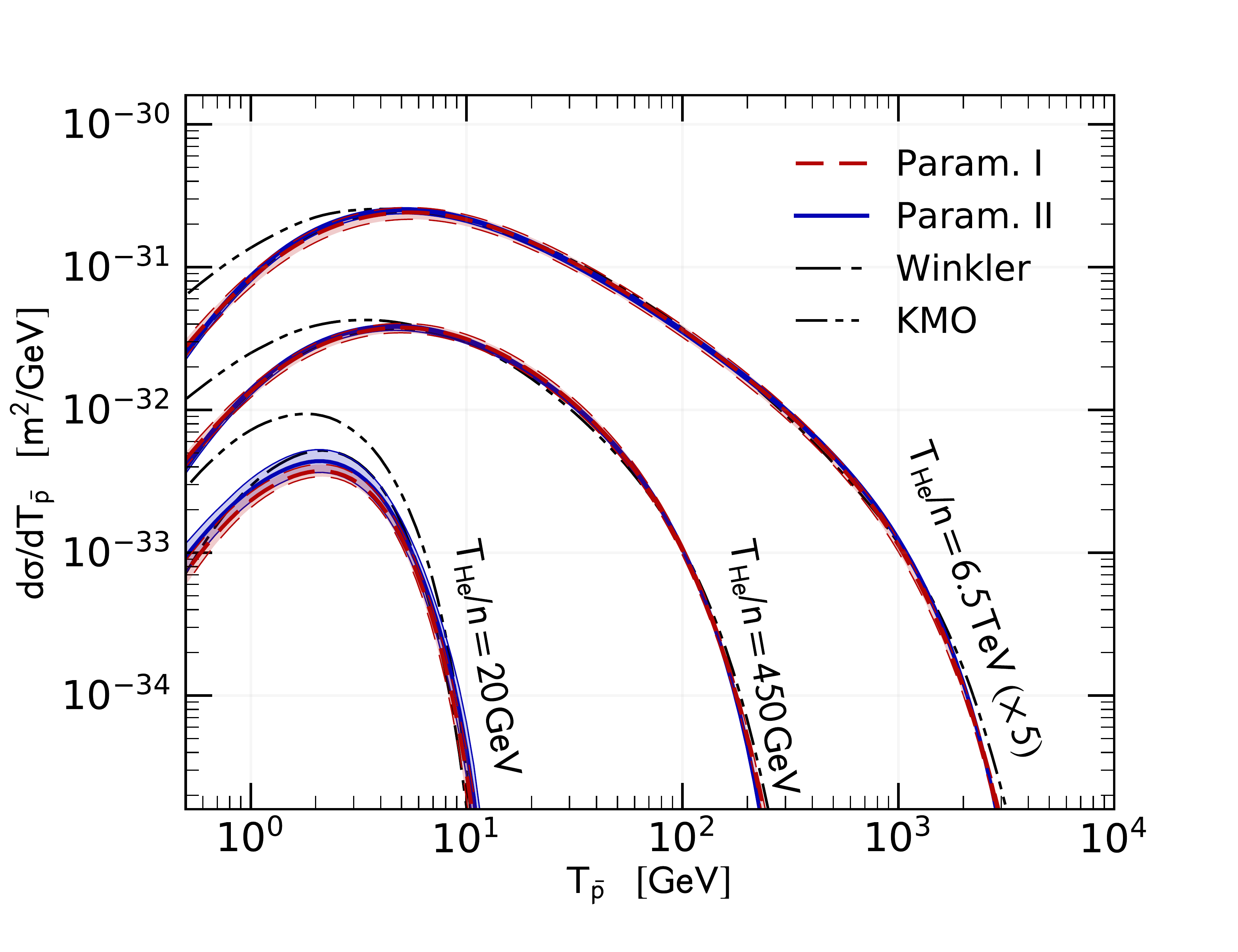}
  \caption{ The differential cross section $d\sigma/dT_{\pbar} (p+{\rm He} \rightarrow \pbar + X)$ (left) and 
  $d\sigma/dT_{\pbar}({\rm He} + p\rightarrow \pbar + X)$ (right) for prompt antiprotons, at the representative incident energies
  $T_p$ = 20 GeV, 450 GeV and 6.5 TeV. The dashed (solid) line and the relevant red (blue) band are the result of our analysis for Param. I and Param. II.   
  The uncertainty band corresponds to the $2\sigma$ confidence interval.
  We report for comparison some literature estimations (see text for details). Tables with the full cross section results are provided in the Supplemental Material to this paper.
           }
	\label{fig::comparison_XS_pHe}
\end{figure*}

\begin{figure*}[t!]
	\includegraphics[width=0.5\textwidth]{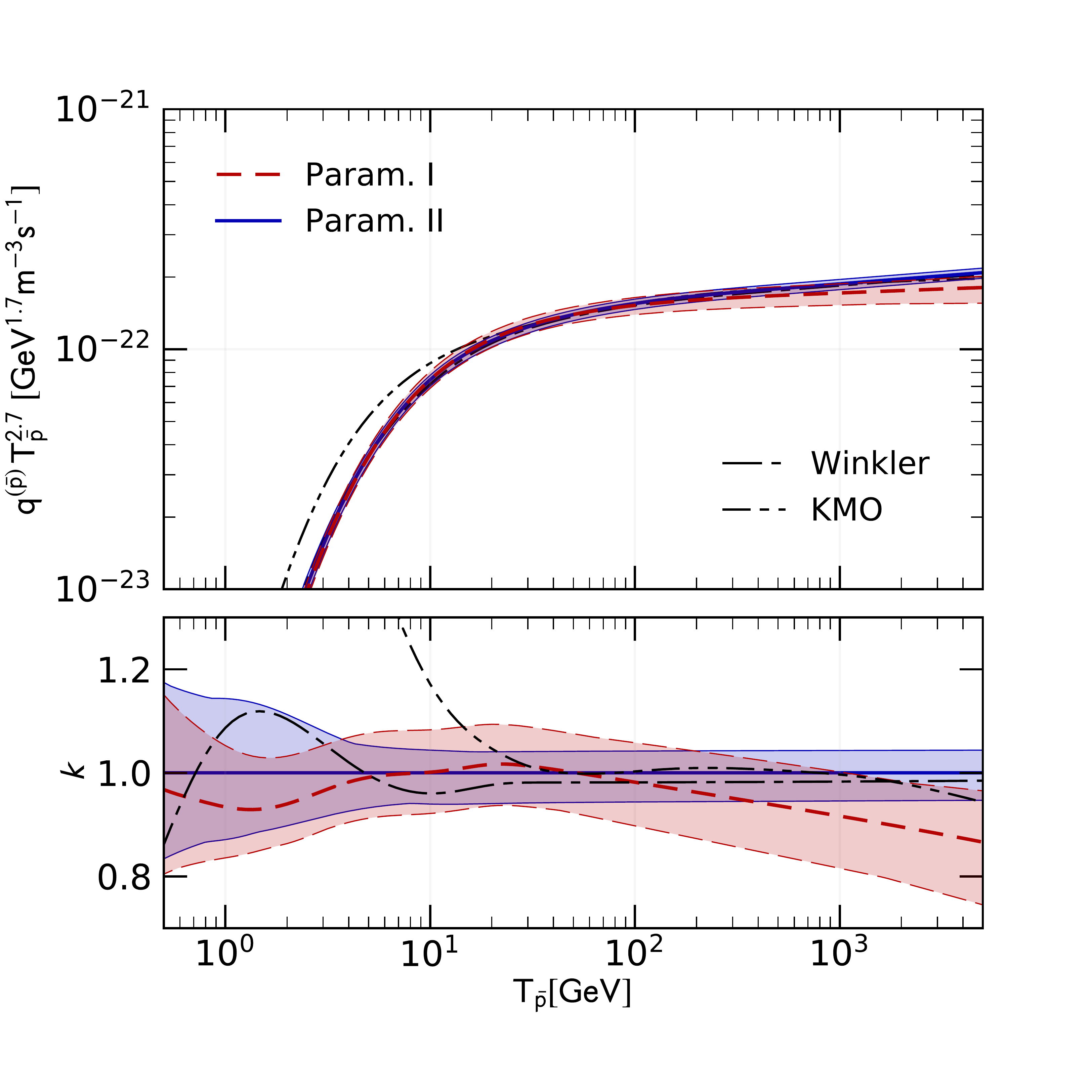}\includegraphics[width=0.5\textwidth]{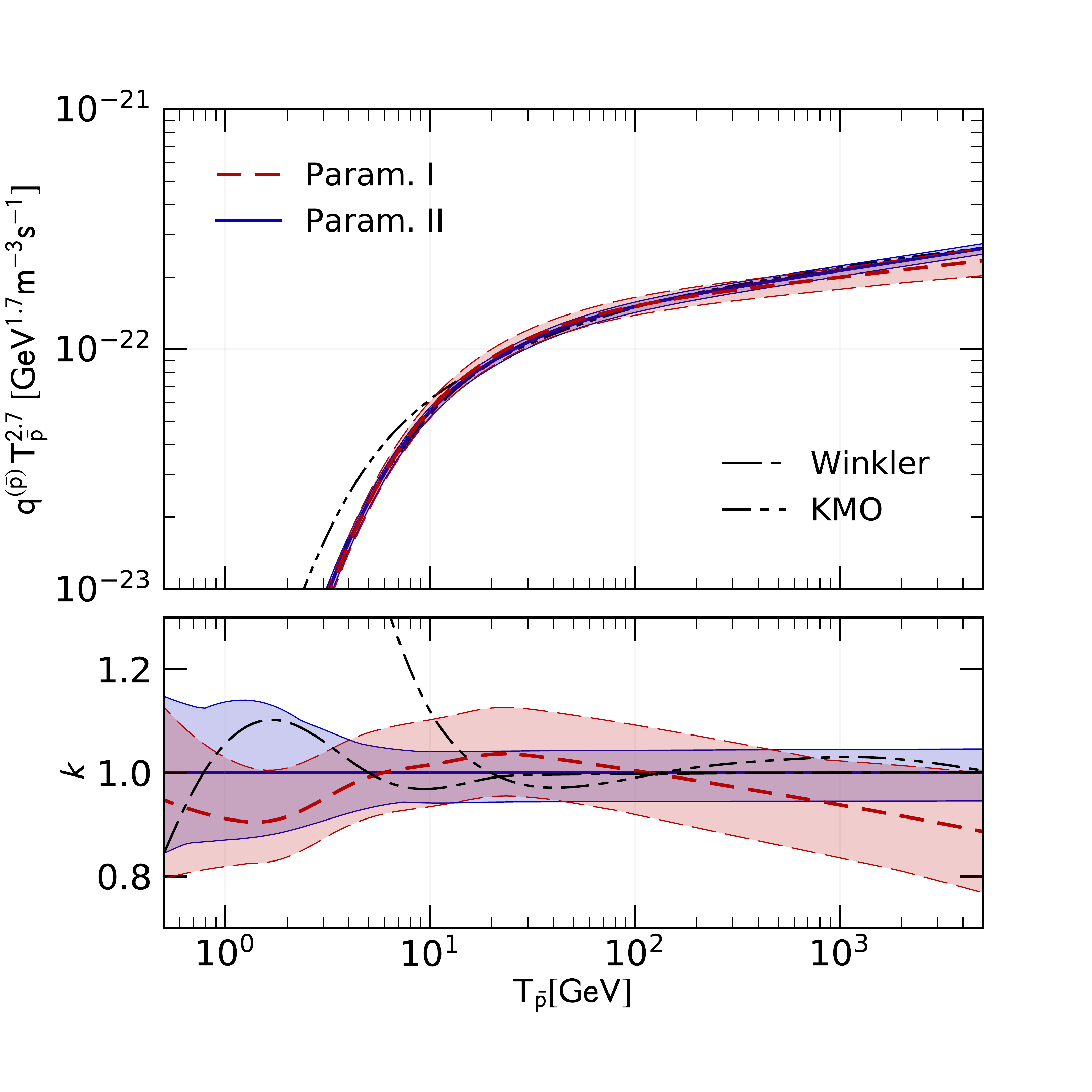}
  \caption{ CR $p$He (left panel) and He$p$ (right panel) antiproton source term with the uncertainty on cross sections 
            for the best fit of Param.~I-B and II-B, {\it i.e.} with NA49 $p$C and LHCb $p$He data. 
            Uncertainties are given at the $2\sigma$ confidence interval.
           }
	\label{fig::comparison_pHe}
\end{figure*}

We perform four fits to determine, first, the goodness of the parametrizations (I~and~II) from the $pp$ fit for the 
interpretation of nuclei data and, secondly, the impact of LHCb data  by excluding (case A) or including (case B) them in the fits. 
\tabref{Fit_properties_pA} comprises the results of all four fits. 
The fits with $p$C data alone (without LHCb data) I-A and II-A converge to a 
$\chi^2/$ndf of 1.3 and 1.1, respectively, leaving the conclusion that the NA49 proton-carbon data fits very well to a rescaled $pp$ cross section.
In the second step, we use the fit results to predict the $p$He cross section and to compare it to LHCb data. 
Param.~I shows a large difference between data and the prediction, measured by a $\chi^2/$ndf from LHCb alone of
9.3. On the other hand, Param.~II gives a $\chi^2/\mathrm{ndf}=1.6$, hinting already the good agreement with Param.~II rescaled by the form factor $f^{pA}$ fixed on $p$C data. 
Including the LHCb data in the fit  does not change the general picture. The quality of the fit slightly improves  to 8.4 and 1.4 in both cases  I-B and II-B, respectively. 
We conclude that Param.~II results in a much better description of the $p$He data by LHCb.
The best fit values for all parameters are summarized in \tabref{Fit_results_pA}. 
Our results for Param.~II agree with \cite{Winkler:2017xor} ($i.e.$  $\langle\nu_\mathrm{He}\rangle=1.25$ there is comparable with 
$4^{D_2}=1.22$ and $4^{1-D1}=1.27$). However, for $p$C we obtain a 10\% lower value of $12^{D_2}=1.43$ or $12^{1-D1}=1.53$ instead
of $\langle\nu_\mathrm{C}\rangle=1.6$.
\figref{LHCb_data} displays the comparison of the LHCb data to the cross section prediction. It is visible that 
the $\pT$-shape of Param.~I does not fit well the data. This shape is inherited from the $pp$ data, and it is therefore  
unlikely to improve the fit by a mere refinement of the  $f^{pA}$ parametrization. We remind that Param.~II includes corrections to
the $\pT$-shape due to scaling violation \cite{Winkler:2017xor}.
Finally, we remark that we explicitly tried a fit solely on LHCb data, but since 
the data contain, apart from one data point, only points for antiprotons produced in backward direction it cannot constrain the asymmetry imposed by $D_2$ and the parameters $D_1$ and $D_2$ turn out to be degenerate.
To calculate $\sigma_{\mathrm{He} + p \rightarrow \pbar +X}$ we use a generalization of \eqnref{param_pA_II}: 
\begin{eqnarray}
  \label{eqn::param_pA_III}
  f^{A_1A_2} &=&  A_1^{D_1} A_2^{D_1} \left[ A_1^{D_2} \left(1+\frac{N_1}{A_1}\Delta_\mathrm{IS} \right) F_\mathrm{pro}(\xF) \right. \qquad \\ \nonumber
         && \left. \qquad  \qquad +   A_2^{D_2} \left(1+\frac{N_2}{A_2}\Delta_\mathrm{IS} \right) F_\mathrm{tar}(\xF) \vphantom{\frac{N}{A}}  \right].
\end{eqnarray}
We cross-checked the validity of this approach by taking the $p$He cross section 
and transforming it to the frame where  the proton is at rest. The two methods give the same result.
Similarly to \figref{comparison_XS_pp}, in \figref{comparison_XS_pHe} we report the results for the differential cross section 
$d\sigma/dT_{\pbar}(p+{\rm He}\rightarrow \pbar + X)$ (left panel) for the representative proton energies $T_p$ = 20 GeV, 450 
GeV and 6.5 TeV. The right panel contains the same information but for incoming He on $p$. 
To determine the fit uncertainty we sample points from the correlation matrices of the $pp$ and $pA$ fits (see \appref{app_corr_matr}). 
To constrain the $pA$ fit at the 2$\sigma$  confidence level
we require that the total $\chi^2$ (sum of $pp$ and $pA$ fit) 
lies within a range of $\Delta\chi^2$=4 compared to our best fit values.
Additionally, we reflect the fact that the $pp$ fit is a prior of the $pA$ fit by requiring that also its 
$\chi^2$ is within a range of $\Delta\chi^2$=4. 
The conclusions are similar to  the 
$pp$ case. We provide a full table for the total cross sections of a number of incident nuclei and their isotopes, and for both $p$ and 
He fixed target in the Supplemental Material to this paper.

We use the fit results to calculate source terms for the $p$He and He$p$ production channels. 
The results are shown in \figref{comparison_pHe}. 
In general, the form and uncertainty of $p$He and He$p$ look similar to $pp$, since both depend on the $pp$ parametrization. 
The fact that He$p$ is harder than $p$He comes from the CR flux which is harder for He compared to $p$.
The two parameterizations are compatible within uncertainties in the AMS-02 $\Tpbar$ energy range, while Param.~I implies a slightly softer $\pbar$ spectrum w.r.t. Param.~II. 
The agreement with former parametrizations Winkler and KMO is unchanged compared to the $pp$ study. However, the re-scaled Di Mauro \etalS shows large deviation in the shape at high energies. We recommend  to use the re-scaling from this paper instead.

\section{\label{sec::final_spectrum} The total antiproton source term}

\begin{figure}[b!]
	\includegraphics[width=.5\textwidth]{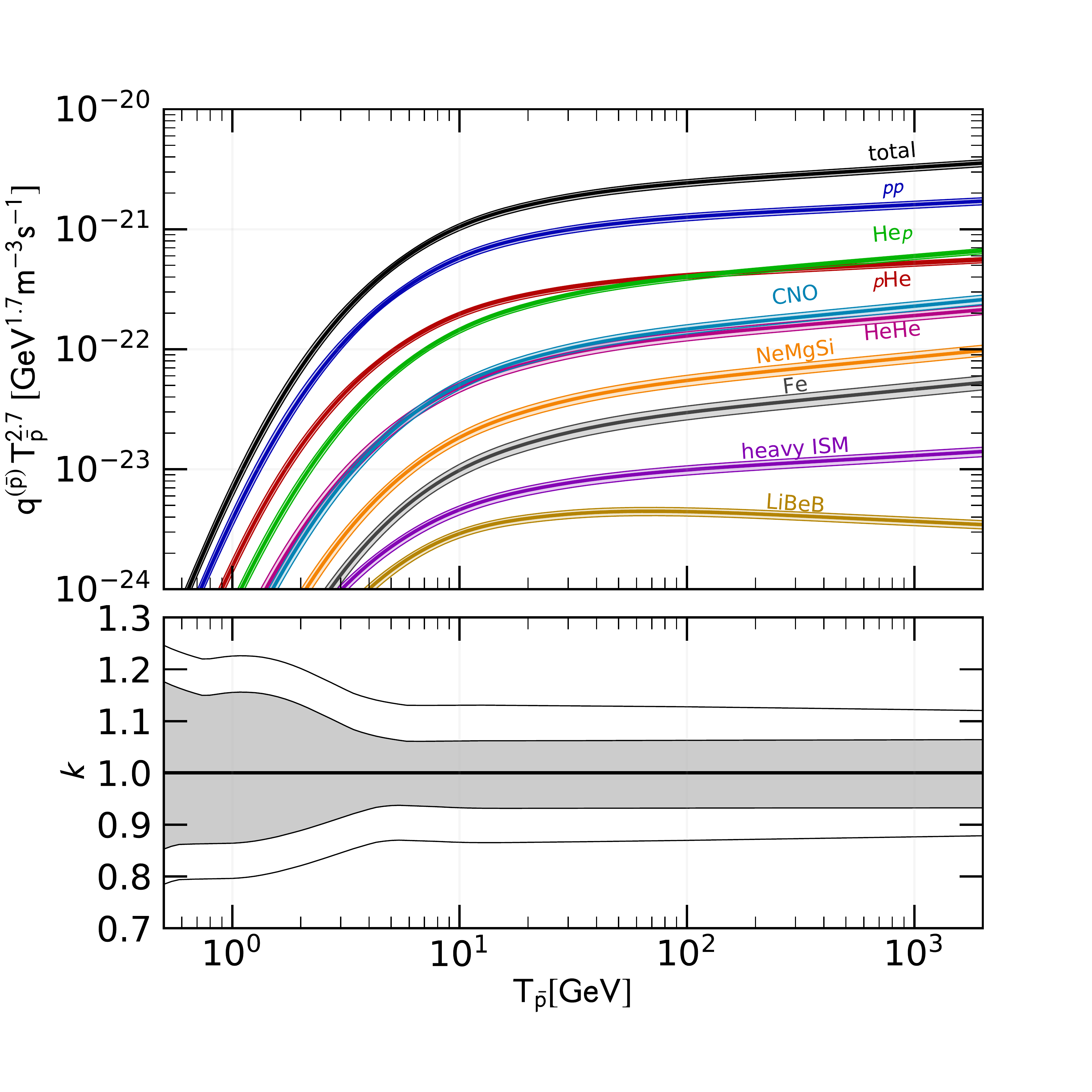}
  \caption{Source terms of CR antiprotons and separate  CR-ISM contributions, grouped  following the prescriptions in \figref{contribution_sourceTerm}. 
  The shaded bands report the $2 \sigma$ uncertainty due to prompt $\pbar$ production cross sections as derived in this paper. 
  In the bottom panel we show the relative uncertainty on the total source term. The grey band refers to the prompt $\pbar$'s 
  only, while the outer lines quantify the additional uncertainty due to isospin violation and to hyperon decay.
   }
	\label{fig::comparison_all}
\end{figure}
The results obtained in the previous sections can be joint to compute the total antiproton source term in the Galaxy, including antineutrons and 
antihyperons, and the contributions from nuclei heavier than helium. The latter, as shown in \figref{contribution_sourceTerm}, give a contribution which is not negligible when compared to errors on the $\pbar$ flux measured by AMS-02. The  CR CNO on $p$ or He contributes 
to the source term at the few percent level each. Even the heavier CR primaries NeMgSi and Fe may contribute above 1\%. 
We note that our fit is tuned to He and C data and therefore the uncertainty on cross sections is extrapolated for CR sources heavier than CNO. 
The total $\pbar$ source term is plotted in \figref{comparison_all}, along with the contribution for every production channel. 
We use the same inputs for CR fluxes and ISM components as discussed in the context of \figref{contribution_sourceTerm}.
It is visible how the measured hardening of CR nuclei fluxes with respect to protons \cite{AMS-02_Aguilar:2015_HeliumFlux,Aguilar:2017hno}
results in a corresponding hardening of the antiproton source term \cite{Donato:2010vm}. 
The rescaling from the prompt $\pbar$ production follows \eqnref{XS_galaxy}. 
We also plot the uncertainty band from the production cross sections, as determined in the fits to data on prompt antiprotons.  
In order to include the production from neutron and hyperon decays, we pick the parameters 
as declared in \cite{Winkler:2017xor}, and namely   
$c_1^\mathrm{IS} = 0.114\pm0.1$  for the determination of $\Delta_{\rm IS}$ (see \eqnref{isospin}), and 
$0.81\pm0.04, c_1^\Lambda = 0.31\pm0.0375$, $c_2^\Lambda = 0.30\pm0.0125$ for the determination of $\Delta_\Lambda$ (see \eqnref{hyperon}). 

The results in \figref{comparison_all} show that the uncertainty due to prompt cross sections (bottom panel) is at the level of $\pm8$\% at 2$\sigma$ above $\Tpbar=5$~GeV. 
At $\Tpbar\leq 5$~GeV it increases to $\pm15$\% at 1~GeV. Adding the uncertainties from  isospin violation in the antineutron production  and from hyperon decays, the uncertainty on the total antiproton source term ranges 
$\pm12$\% from high energies down to about few GeV, and increases to $\pm20$\% below that value.
Above $\Tpbar=50$~GeV the total antiproton source spectrum can be approximated by  a power law  with an index of about $-2.5$ .

\section{\label{sec::happy} Conclusions} 

The role of high-energy particle physics in the interpretation of CR data receives increasing attention, 
since data from space are provided with improving precision.
AMS-02 on the International Space Station collected data of CR nuclei, leptons, and antiprotons with unprecedented accuracy, 
often pushing uncertainties down to few percent in a large range of energy from the GeV to the TeV scale.
The fluxes of secondary CRs, which are produced in interactions with the ISM, depend on the inclusive production 
cross sections provided by high-energy particle experiments. In particular, this applies to CR antiprotons whose origin is believed
to be dominantly secondary. 
Consequently, the interpretation of the antiproton flux in terms of CR propagation or the search for a possible primary component, such as 
for example dark matter annihilation or decay, relies on the accurate modeling of secondary production. The underlying cross sections
should be provided at least at  the same accuracy level as CR measurements.

In this paper,  we analyze the first-ever data on the inclusive cross section $p + \mathrm{He} \rightarrow \pbar +X$ 
collected by the LHCb collaboration at Cern, with beam protons at $\Tp=6.5$~TeV and a fixed helium target. 
Since the coverage of the kinematic parameter space of this data do not allow a standalone parametrization, 
we apply a rescaling from $p + p \rightarrow \pbar +X$ cross section.
Therefore, we update the most recent parametrizations from Di Mauro \etalS (Param.~I) and Winkler (Param.~II) exploiting the newly available NA61 data.
Then we determine the rescaling factor to proton-nucleus using $p$He data from LHCb and $p$C data from NA49 (taken at $\sS = 110$ and 17.3~GeV, respectively). 
The LHCb $p$He data clearly prefer Param.~II. All other data result in equally good fits for both parametrizations. 
Moreover, the LHCb data show for the first time how well the rescaling from the $pp$ channel applies to helium target. By using $pp$, $p$He and $p$C data we estimate the uncertainty on the 
Lorentz invariant cross section for $p + \mathrm{He} \rightarrow \pbar + X$.
This uncertainty is dominated by $p+p \rightarrow \pbar + X$ cross section, which translates into all channels 
since we derive them using the $pp$ cross sections.

Finally, we use our cross sections to compute the antiproton source terms and their uncertainties for all the production channels, considering also nuclei heavier than He both in CRs and the ISM. 
At intermediate energies from $\Tpbar=5$~GeV up to a few hundred GeV the prompt source terms derived from Param.~I and II are 
compatible within uncertainties, which are at the level of $\pm8$\% at the 2$\sigma$ level and increase to $\pm15$\% below $\Tpbar=5$~GeV. 
The uncertainty is dominated by $p+p \rightarrow \pbar + X$ cross section, which translates into all channels. 
Antineutron- and  hyperon-induced production increases the uncertainty by an additional 5\%.
Overall the secondary antiproton source spectrum is affected by an uncertainty of up to $\pm20$\%.
Moreover, we find that 
CR CNO makes up to few percent of the total source term and should always be considered.
In the Supplemental Material to this paper, we provide the energy-differential cross sections, which are required to calculate the source spectrum, for all relevant isotopes. 
We quantify the necessity of new data on antiproton production cross sections, and pin down the kinematic parameter space 
which should be covered by future data.

\section{\label{sec::acknowledgments}Acknowledgments}
We warmly thank G. Graziani and G. Passaleva for useful discussions, and P. Von Doetinchem for many useful insights on NA61 data.

\bibliographystyle{apsrev4-1.bst}
\bibliography{bibliography}{}

\appendix
\section{Correlation matrices.}
\label{app::app_corr_matr}

\begin{figure*}[t!]
	\includegraphics[height=5.3cm, trim={1.1cm 0 3cm 0},clip]{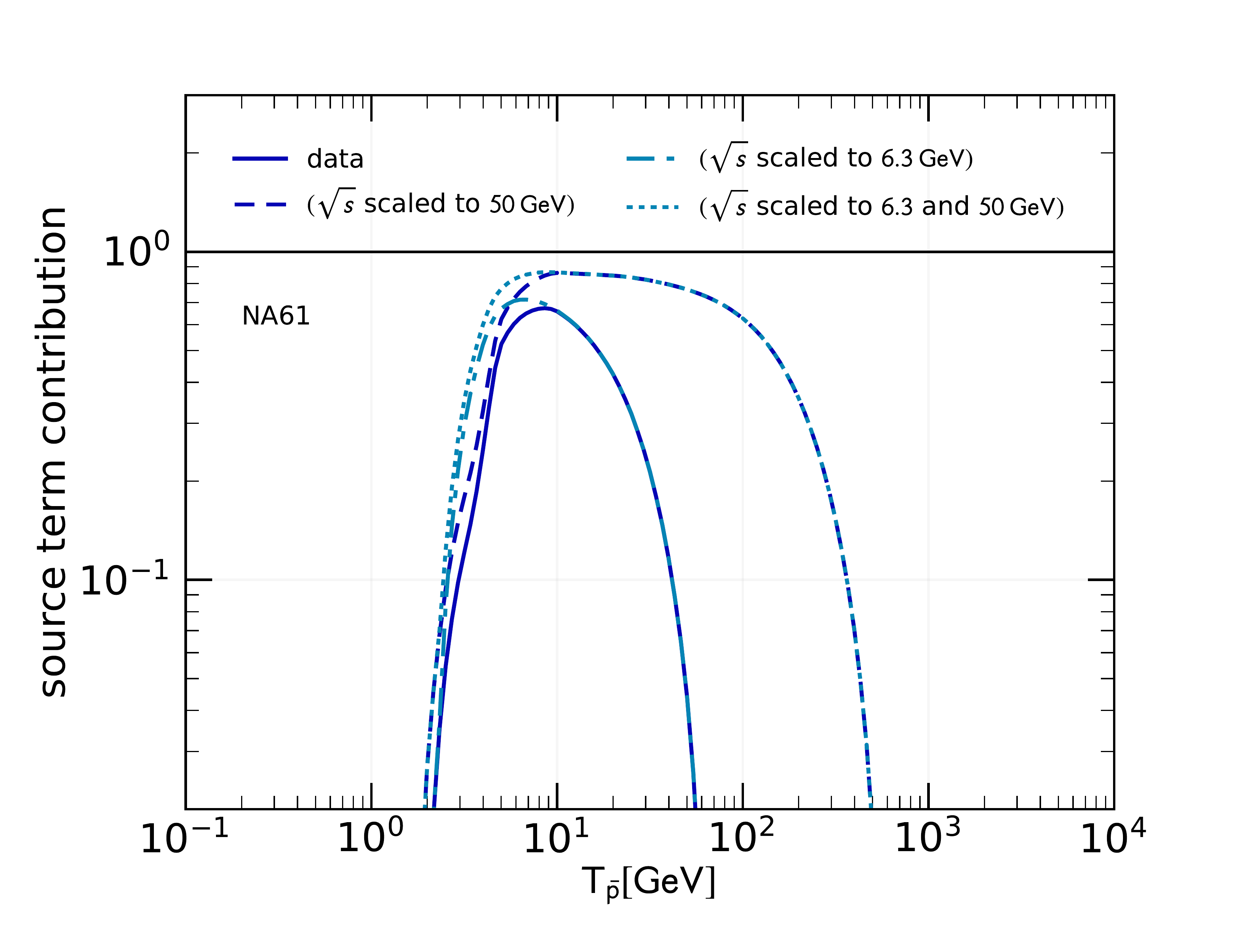} \includegraphics[height=5.3cm, trim={2.2cm 0 3cm 0},clip]{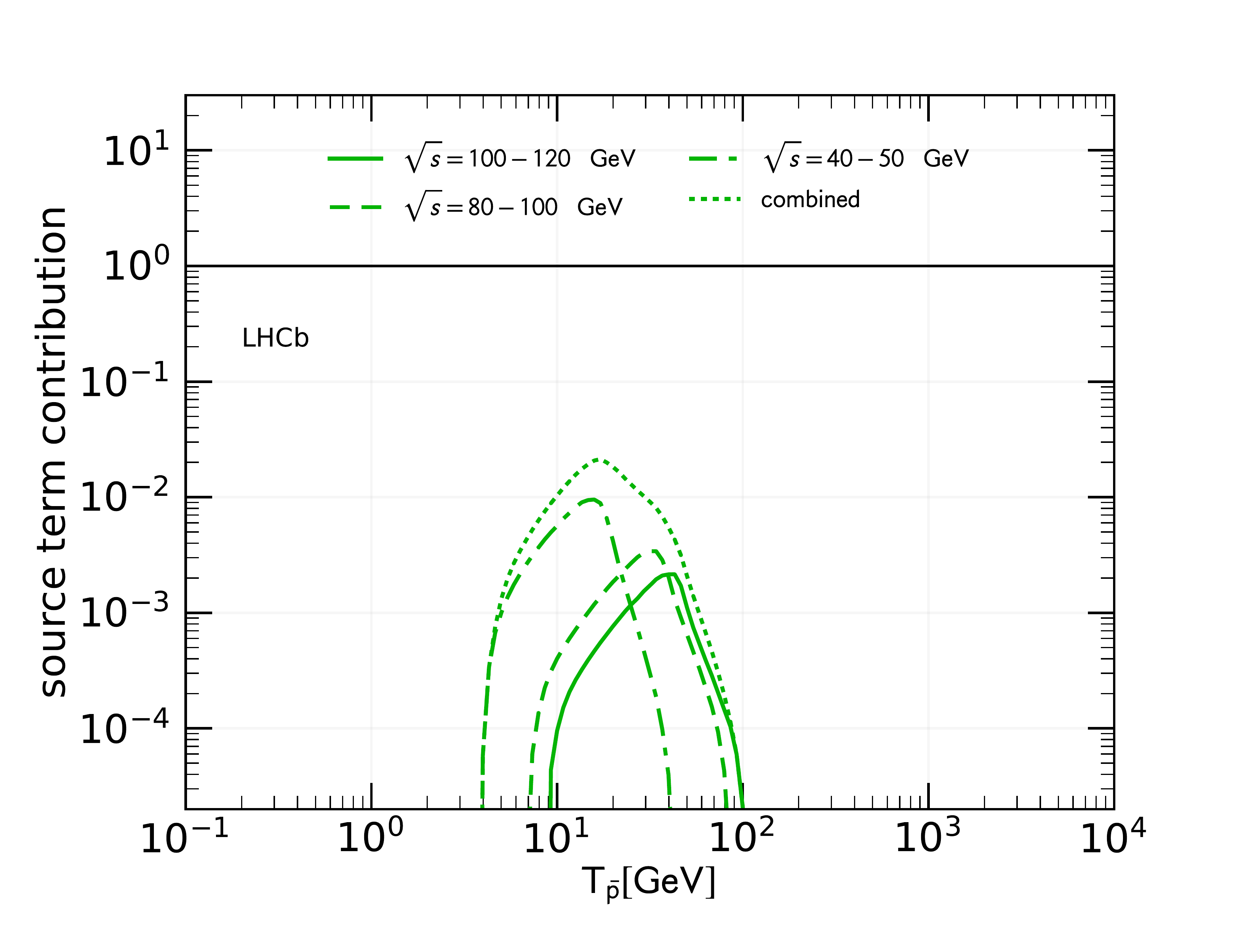}\includegraphics[height=5.3cm, trim={2.2cm 0 3cm 0},clip]{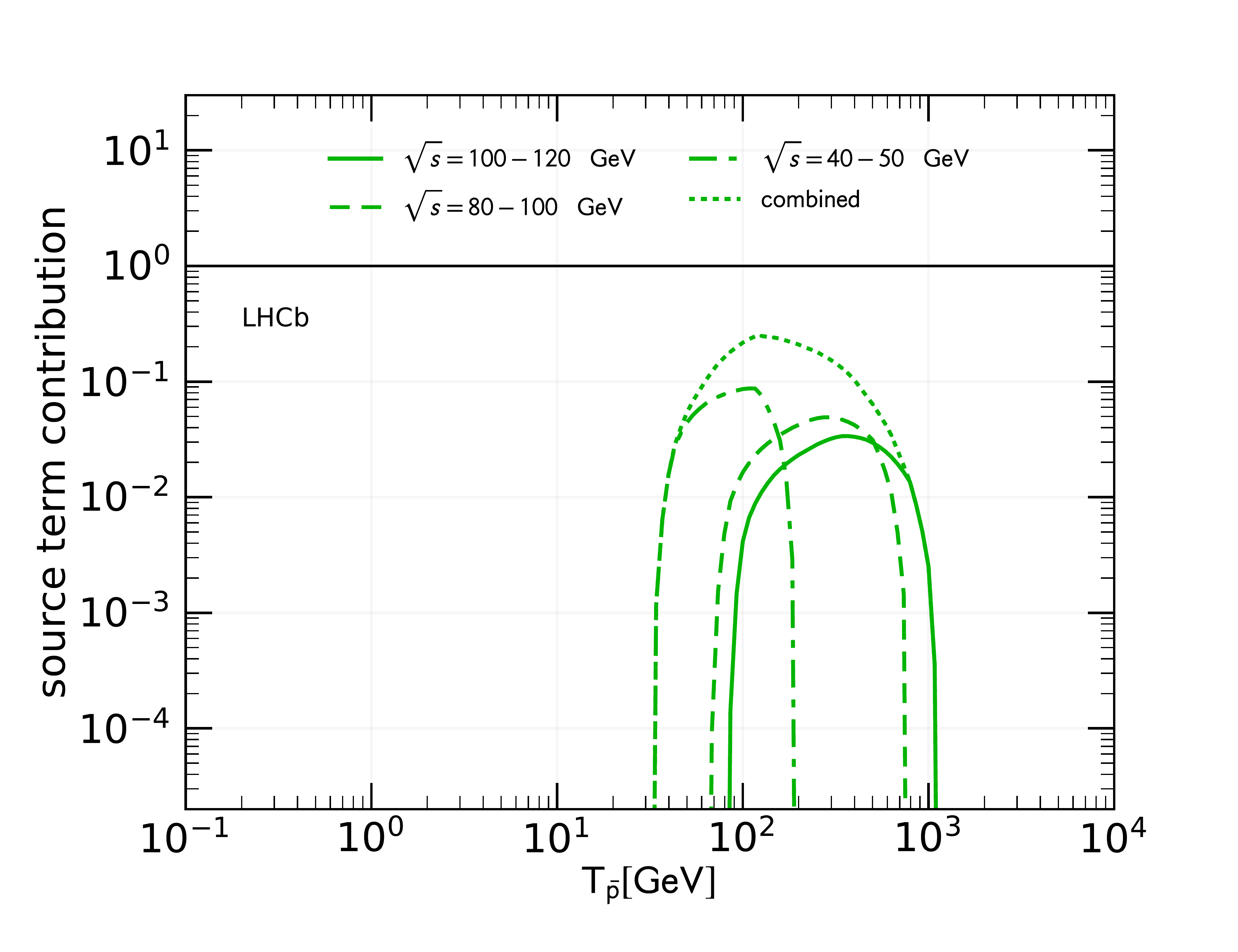}
	\caption{ 	Similar to \figref{contribution_pp} and \figref{contribution_pHe}.           
           			 Fraction of the antiproton source term originating from the kinematic parameter space of the cross section 
           			 which currently is  experimentally determined by NA61 data in the $pp$ channel (left panel) and by 
           			 LHCb data in the  $p$He (central panel) or He$p$ (right panel)  channels. We add future predictions for a
           			 possible evaluation of NA61 data at $\sS=6.3$~GeV and LHCb measurements at $\sS=43$ and 87~GeV. 
           			 Each contribution is normalized to the total source term of the specific  channel.	 
				}
	\label{fig::contribution_future}
\end{figure*}

\begin{figure*}[t!]
	\includegraphics[width=0.5\textwidth]{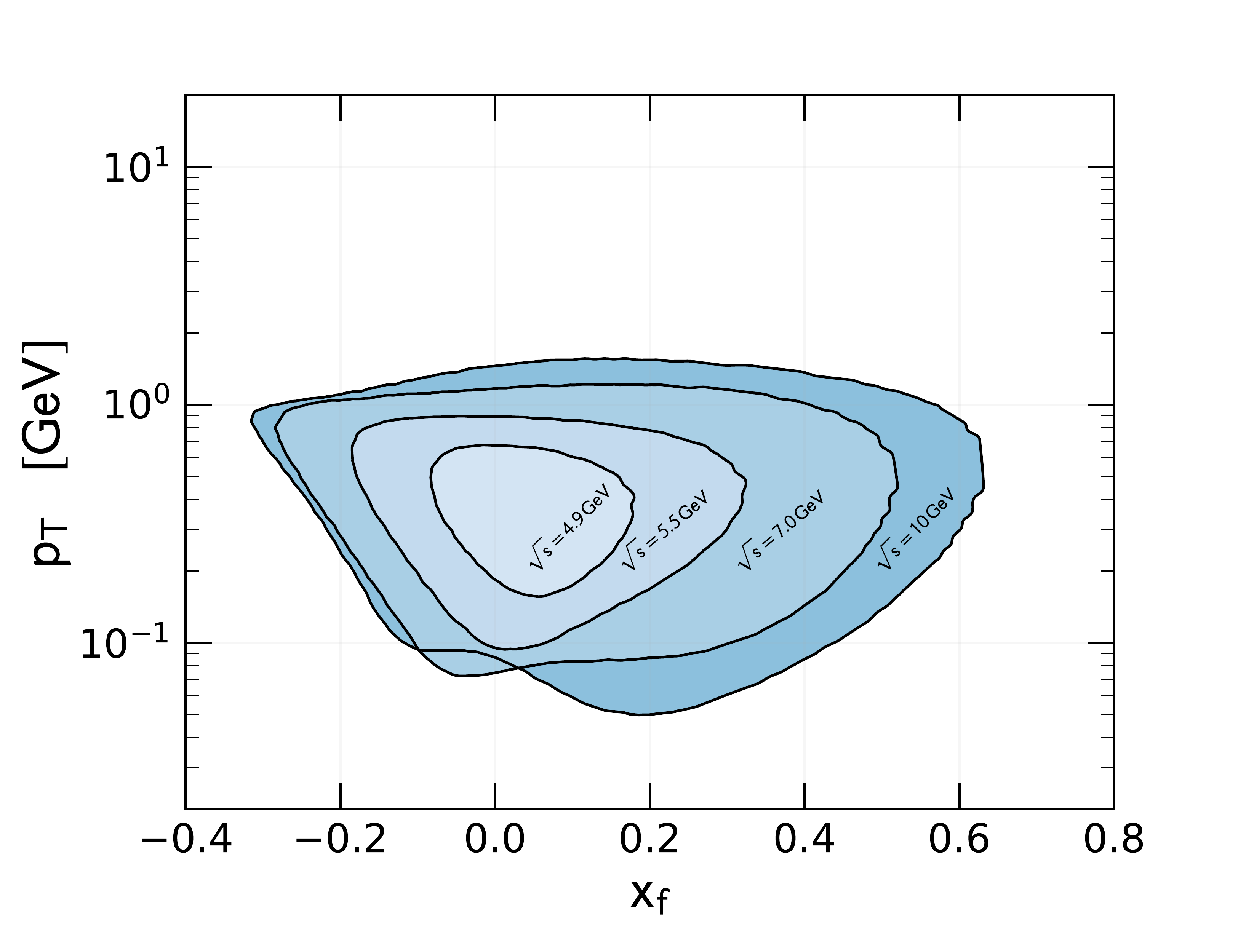}\includegraphics[width=0.5\textwidth]{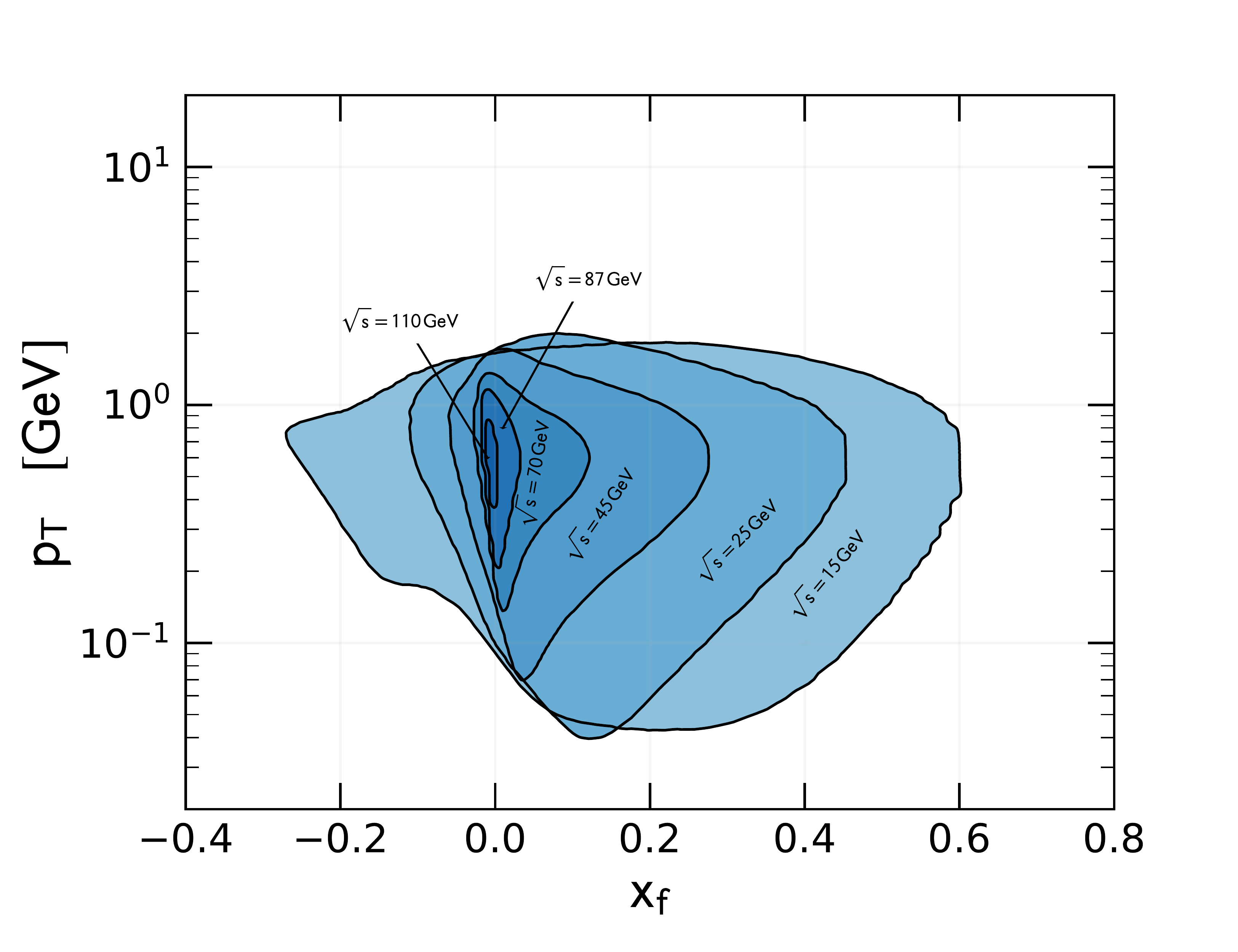}
	\caption{ 	Parameter space of the  antiproton production cross section which is necessary to determine 
	         		the antiproton source term at the uncertainty level of AMS-02 measurements 
	        		 	\cite{AMS-02_Aguilar:2016_AntiprotonFlux}. 
	         		We require  the cross section to be known by 3\% within the blue shaded regions
	         		and by 30\% outside of the contours. The left and right panels contain contours for 
	         		different CM energies.
	         		This figure is an update of Fig.~7b in DKD17. We exchange the kinetic variable $\xR$ by $\xF$, which
	         		is suitable for the asymmetric $pA$ cross section discussed in this paper. 
	}
	\label{fig::pbar_measurment_xF}
\end{figure*}

In \tabref{Correlation_pp_diMauro} and \tabref{Correlation_pp_Winkler} we report the correlation matrices of 
the fits performed on the $pp$ data with Param.~I and~II, respectively. Equivalently, 
\tabref{Correlation_pA_diMauro} and \tabref{Correlation_pA_Winkler} contain the correlation matrices of 
the $pA$ form factor Param.~I-B and~II-B.
\begin{table*}[h]
\caption{Correlation matrix of the $pp$ fit with Param.~I.}
\label{tab::Correlation_pp_diMauro}
\begin{tabular}{ l | r r r r r r r r r r r r | }
      & $C_1$     & $C_2$     & $C_3$     & $C_4$     & $C_5$     & $C_6$     & $C_7$     & $C_8$     
      & $\omega_\mathrm{BRAHMS}$
      & $\omega_\mathrm{Dekkers}$
      & $\omega_\mathrm{NA49}$
      & $\omega_\mathrm{NA61}$
      \\ \hline 
$C_1$ & 1.000     &-0.994     &0.000     &-0.077     &-0.002     &-0.107     &-0.019     & 0.131     &-0.057     &-0.226     & 0.100     & 0.117        \\
$C_2$ &-0.994     & 1.000     & 0.017     & 0.071     &0.000     & 0.152     &-0.015     &-0.153     & 0.042     & 0.208     &-0.082     &-0.114        \\
$C_3$ &0.000     & 0.017     & 1.000     &-0.847     &-0.488     &-0.217     & 0.288     & 0.452     &-0.155     & 0.160     & 0.140     & 0.155        \\
$C_4$ &-0.077     & 0.071     &-0.847     & 1.000     & 0.810     & 0.246     &-0.355     &-0.647     & 0.234     &-0.022     & 0.007     &-0.015        \\
$C_5$ &-0.002     &0.000     &-0.488     & 0.810     & 1.000     & 0.272     &-0.450     &-0.750     &-0.026     & 0.088     & 0.011     & 0.051        \\
$C_6$ &-0.107     & 0.152     &-0.217     & 0.246     & 0.272     & 1.000     &-0.923     &-0.422     &-0.417     & 0.228     & 0.400     & 0.457        \\
$C_7$ &-0.019     &-0.015     & 0.288     &-0.355     &-0.450     &-0.923     & 1.000     & 0.532     & 0.541     &-0.157     &-0.131     &-0.249        \\
$C_8$ & 0.131     &-0.153     & 0.452     &-0.647     &-0.750     &-0.422     & 0.532     & 1.000     & 0.042     &-0.039     &-0.029     &-0.058        \\
$\omega_\mathrm{BRAHMS}$  &-0.057     & 0.042     &-0.155     & 0.234     &-0.026     &-0.417     & 0.541     & 0.042     & 1.000     &-0.020     & 0.302     & 0.153        \\
$\omega_\mathrm{Dekkers}$ &-0.226     & 0.208     & 0.160     &-0.022     & 0.088     & 0.228     &-0.157     &-0.039     &-0.020     & 1.000     & 0.325     & 0.371        \\
$\omega_\mathrm{NA49}$    & 0.100     &-0.082     & 0.140     & 0.007     & 0.011     & 0.400     &-0.131     &-0.029     & 0.302     & 0.325     & 1.000     & 0.894        \\
$\omega_\mathrm{NA61}$    & 0.117     &-0.114     & 0.155     &-0.015     & 0.051     & 0.457     &-0.249     &-0.058     & 0.153     & 0.371     & 0.894     & 1.000     
\end{tabular}
\end{table*}

\begin{table*}[h]
\caption{Correlation matrix of the $pp$ fit with Param.~II. Note that $C_4$ is a fixed parameter and therefore uncorrelated to the other parameters.}
\label{tab::Correlation_pp_Winkler}
\begin{tabular}{ l | r r r r r r r r r| }
      & $C_1$     & $C_2$     & $C_3$     & $C_4$     & $C_5$     & $C_6$
      & $\omega_\mathrm{Dekkers}$
      & $\omega_\mathrm{NA49}$
      & $\omega_\mathrm{NA61}$
      \\ \hline 
$C_1$ & 1.000     & 0.338     & 0.003     & 0.000     &-0.214     & 0.055     & 0.057     & 0.951     & 0.843     \\
$C_2$ & 0.338     & 1.000     & 0.312     & 0.000     & 0.207     & 0.355     & 0.035     & 0.075     &-0.092     \\
$C_3$ & 0.003     & 0.312     & 1.000     & 0.000     & 0.097     & 0.106     & 0.005     & 0.017     &-0.019     \\
$C_4$ & 0.000     & 0.000     & 0.000     & 1.000     & 0.000     & 0.000     & 0.000     & 0.000     & 0.000     \\
$C_5$ &-0.214     & 0.207     & 0.097     & 0.000     & 1.000     &-0.127     & 0.666     &-0.282     &-0.289     \\
$C_6$ & 0.055     & 0.355     & 0.106     & 0.000     &-0.127     & 1.000     & 0.117     &-0.043     &-0.056     \\
$\omega_\mathrm{Dekkers}$ & 0.057     & 0.035     & 0.005     & 0.000     & 0.666     & 0.117     & 1.000     & 0.050     & 0.065     \\
$\omega_\mathrm{NA49}$    & 0.951     & 0.075     & 0.017     & 0.000     &-0.282     &-0.043     & 0.050     & 1.000     & 0.919     \\
$\omega_\mathrm{NA61}$    & 0.843     &-0.092     &-0.019     & 0.000     &-0.289     &-0.056     & 0.065     & 0.919     & 1.000
\end{tabular}
\end{table*}

\begin{table*}[h]

\begin{minipage}{.5\linewidth}
\caption{Correlation matrix Param. I-B.}
\label{tab::Correlation_pA_diMauro}
\begin{tabular}{ l | r r r r | }
      & $D_1$     & $D_2$     
      & $\omega_\mathrm{LHCb}$
      & $\omega_\mathrm{NA49}$
      \\ \hline 
$D_1$ & 1.000     &-0.516     & 0.603     & 0.798     \\
$D_2$ &-0.516     & 1.000     & 0.169     &-0.008     \\
$\omega_\mathrm{LHCb}$  & 0.603     & 0.169     & 1.000     & 0.745      \\
$\omega_\mathrm{NA49}$  & 0.798     &-0.008     & 0.745     & 1.000     \\
\end{tabular}
\end{minipage}\begin{minipage}{.5\linewidth}
\caption{Correlation matrix Param. II-B.}
\label{tab::Correlation_pA_Winkler}
\begin{tabular}{ l | r r r r | }
      & $D_1$     & $D_2$     
      & $\omega_\mathrm{LHCb}$
      & $\omega_\mathrm{NA49}$
      \\ \hline 
$D_1$ & 1.000     &-0.496     & 0.598     & 0.813     \\
$D_2$ &-0.496     & 1.000     & 0.216     &-0.017     \\
$\omega_\mathrm{LHCb}$  & 0.598     & 0.216     & 1.000     & 0.749      \\
$\omega_\mathrm{NA49}$  & 0.813     &-0.017     & 0.749     & 1.000     \\
\end{tabular}
\end{minipage}

\end{table*}

\section{Source term fraction in the future }
\label{app::app_source_term_fraction}

The derivation of the source term in this paper reveals that the uncertainty of the cross sections to calculate the source term of CR antiprotons 
is still large compared uncertainties in the antiproton flux measured by AMS-02. This is partially due to the fact that the cross section coverage of the source term, namely, the fraction of the source term determined by the parameter space of cross section experiments is relatively small. 
In the context of \figref{contribution_pp} we discuss the situation of the $pp$ channel. One very important step is to improve the coverage at low energies. 
NA61 has taken data of $pp$ collisions at $\sS=6.3$, 7.7, 8.8, 12.3, and 17.3~GeV, but 
evaluated $p + p \rightarrow \pbar + X$ only from $\sS=7.7$~GeV.
In \figref{contribution_future} (left panel) we show that the coverage of the source term could be improved down to $\Tpbar=3$~GeV if 
NA61 would be able to analyze this data for antiprotons. We assume that the coverage in $\xR$ and $\pT$ is comparable to the measurement at $\sS=7.7$~GeV. 

Similarly, one can guess further potentials in the $p$He channels.  
The LHCb data are taken at very high energies of $\sS=110$~GeV and, therefore, their antiproton production in the energy range interesting for CRs results in a very small contribution to the source term, as shown in \figref{contribution_pHe}.  
We estimate the fraction of the $\pbar$ source term  for measurements at $\sS=43$ and 87~GeV,  where we assume equal coverage in $\xF$ and $\pT$ as 
for the LHCb data at  $\sS=110$~GeV. In \figref{contribution_future} we show the source term fraction these measurements could achieve in the $p$He (central panel)
and He$p$ (right panel) channel. These measurements and especially their combination would significantly improve the coverage of the helium channels by LHCb.

\section{Parameter space explorability}
\label{app::app_par_space}

In DKD17 we studied the precision of cross section measurements which would be necessary to shrink the uncertainties imposed on the theoretical prediction of the antiproton flux such that they are on the same level as flux measurement by AMS-02. We identified the relevant parameter space to be covered by high-energy particle physics experiments. We represented our results both in the LAB frame, where the target is at rest, and in the CM frame. Since we focused on proton-proton scattering and assumed a symmetric cross section, we presented our results for the CM system in terms of $\xR$. In this paper we discussed in detail the asymmetry in proton-nucleus scattering. The appropriate equivalent to $\xR$ in this case is $\xF$. Therefore, here we update our main result from DKD17 and present the parameter space which should be covered by experiments in terms of $\xF$. The results presented in \figref{pbar_measurment_xF}  are now applicable to any proton-nucleus production channel.

\end{document}